\documentclass[12pt]{spieman}  
\usepackage{amsmath,amsfonts,amssymb}
\usepackage{graphicx}
\usepackage{setspace}
\usepackage{tocloft}
\usepackage{lineno}

\usepackage[skip=5pt, indent=14pt]{parskip}

\newcommand{\CMtable}{\fontfamily{cmr}\selectfont}
\usepackage{booktabs}
\renewcommand{\arraystretch}{1.25} 

\title{Predicted Capabilities of the SPRITE SmallSat for a Low-Redshift Lyman Continuum Emission Survey}
\author[a,*]{Yi Hang Valerie Wong}
\author[a]{Brian Fleming}
\author[a]{Elena Carlson}
\author[a]{Briana Indahl}
\author[a]{Dmitry Vorobiev}
\author[a]{Sebastian Escobar}
\author[a]{Kevin France}
\author[a]{Maitland Bowen}
\author[a]{Dónal O'Sullivan}
\author[b]{Anne E. Jaskot}
\author[c]{Jason Tumlinson}
\author[d]{Sanchayeeta Borthakur}
\author[e]{Michael J. Rutkowski}
\author[f]{Stephan McCandliss}
\author[c]{Ravi Sankrit}
\author[g]{John O'Meara}

\affil[a]{Laboratory for Atmospheric and Space Physics, University of Colorado Boulder, 1234 Innovation Drive, Boulder, CO 80303, USA}
\affil[b]{Department of Astronomy, Williams College, Williamstown, MA 01267, USA}
\affil[c]{Space Telescope Science Institute, 3700 San Martin Dr., Baltimore MD 21218, USA}
\affil[d]{School of Earth and Space Exploration, Arizona State University, 781 Terrace Mall, Tempe, AZ 85287, USA}
\affil[e]{Minnesota State University-Mankato, Department of Physics \& Astronomy, Trafton Science Center North 141, Mankato, MN 56001, USA}
\affil[f]{Johns Hopkins University, Department of Physics \& Astronomy, Center for Astrophysical Sciences, 3400 North Charles Street, Baltimore, MD 21218, USA}
\affil[g]{W. M. Keck Observatory, 65-1120 Mamalahoa Hwy, Kamuela, HI 96743, USA}

\cftpagenumbersoff{figure}
\cftpagenumbersoff{table} 
\begin{document} 
\maketitle

\begin{abstract}

Ionizing Lyman continuum (LyC; $\lambda <$ 912~\AA) radiation from low-redshift ($z \sim$ 0.3) galaxies provides crucial insight into the processes that contributed to cosmic reionization.
While the \textit{James Webb Space Telescope} has observed galaxies at redshifts as high as $z \sim$ 14, detecting LyC beyond $z \sim$ 3 is challenging due to absorption by neutral hydrogen in the intergalactic medium (IGM).
Low-redshift LyC emitters (LCEs), therefore, act as proxies for their high-redshift counterparts, enabling direct measurements of LyC escape fractions with reduced IGM interference.
These observations allow detailed ancillary studies of galaxy properties and the mechanisms driving ionizing photon escape, which cannot be directly observed at the Epoch of Reionization.
This paper examines the capabilities of the Supernova remnants and Proxies for Re-Ionization Testbed Experiment (SPRITE) SmallSat, designed to study LyC emission from star-forming galaxies at 0.16 $< z <$ 0.4.
SPRITE uses advanced mirror coatings and a highly sensitive far-ultraviolet imaging spectrograph, enabling it to probe LyC from galaxies that have been difficult to study with prior and existing instruments.
To assess SPRITE's predicted performance in LyC studies, we select eight previously confirmed LCEs from the Low-redshift Lyman Continuum Survey as commissioning targets.
Observations of these commissioning LCEs will validate SPRITE’s LyC sensitivity and characterize its detection limits.
This will enable the broader SPRITE low-redshift LCE survey, which will provide new constraints on the physics of LyC escape and help bridge the gap between low- and high-redshift LyC studies.
SPRITE will also inform the design and scientific potential of future Lyman-UV missions, including the Habitable Worlds Observatory.

\end{abstract}


{\noindent \footnotesize\textbf{*}Yi Hang Valerie Wong,  \linkable{valerie.wong@lasp.colorado.edu} }

\begin{spacing}{1}   

\section{Introduction}
\label{sect:intro}  

Recent discoveries with the \textit{James Webb Space Telescope} (\textit{JWST}) have unveiled galaxies with redshifts as high as $z \sim$ 14 [\citenum{2024Natur_633_318I,2023NatAs_7_611R, 2023ApJ_957L_34W}],
where the UV spectrum is redshifted into \textit{JWST}'s infrared bandpass.
These high-redshift galaxies may be the sources of the radiation that led to the nearly complete reionization of the intergalactic medium (IGM) near a redshift of $z \sim$~6, thus ushering in the Epoch of Reionization (EoR).
However, \textit{JWST} is unable to study the morphology or stellar populations in significant detail for such distant targets, nor can it detect the ionizing radiation directly because of absorption by neutral hydrogen (H\text{\ I}) in the line of sight beyond redshift $z \sim$~3.
This ionizing spectrum is known as the Lyman continuum (LyC), defined as photons with rest wavelengths $\lambda \le$~912~\AA, corresponding to the ionization threshold of neutral hydrogen ($E_{\rm{\gamma}} \geq$~13.6~eV).
In practice, LyC measurements usually focus on the wavelength range of $\sim$~700--912~\AA, where LyC escape can be studied while capturing the majority of the escaping ionizing photon budget.

The mean transmission of the IGM just below 912~\AA\ is $\sim$~55\%\ for sources at $z \sim$~3 and decreases to $\sim$~20\%\ at $z \sim$~4 [\citenum{2008MNRAS_387_1681I, JATIS_11_4_042223}].
This dramatic decrease is due to the increasing density of H\text{\ I} in the IGM, higher column densities of H\text{\ I}, and a more significant number of intergalactic absorbers such as the Ly$\alpha$ forest, Lyman limit systems, and damped Ly$\alpha$ systems at higher $z$.
Therefore, it is necessary to study galaxies at redshifts $z < 4$ to allow for the direct measurement of the LyC escape fraction without interference from H\text{\ I} in the IGM.
Observations of galaxies at low redshift also enable us to study the properties of the galaxies in greater detail due to their larger angular size and accessibility of other key diagnostics (e.g., nebular emission-line ratios, UV continuum slope). While these sources may not be the actual galaxies that reionized the universe at 6~$< z <$~14, they provide an essential laboratory for understanding the physics that may have driven this important moment in cosmic history.

Unfortunately, searches for low-redshift LyC sources yielded very few detections for several decades.
First attempts to observe the far-ultraviolet spectra of four starburst galaxies at $z \sim$ 0.02--0.03 were made with the \textit{Hopkins Ultraviolet Telescope} (\textit{HUT}) [\citenum{1995ApJ_454L_19L}].
However, with fluxes close to the noise level, they concluded that there was no detection of LyC in the four galaxies, with 2$\sigma$ upper limits of $F_{\lambda} <$~7~$\times$~10$^{-16}$~ergs~s$^{-1}$ cm$^{-2}$~\AA$^{-1}$.
Up until 2014, the only detections were of Haro 11 at $z\sim$~0.02 and Tol 1247--232 at $z \sim$ 0.04, using the \textit{Far Ultraviolet Spectroscopic Explorer} (\textit{FUSE}) [\citenum{2006AA_448_513B, 2011AA_532A_107L, 2013AA_553A_106L}].
Despite the high sensitivity of the wavelength coverage for redshifted Lyman break in galaxies at $z <$~0.3, these are the only two galaxies in the \textit{FUSE} archive with detected LyC.

A milestone in low-redshift LyC detection was achieved in 2016 when five Green Pea galaxies were observed with Cosmic Origins Spectrograph on the \textit{Hubble Space Telescope} (\textit{HST}-COS) at $z \sim$~0.3.
The five galaxies were selected from the Sloan Digital Sky Survey (SDSS) with the following criteria:
(1) compact, non-spiral galaxies without active galactic nuclei (AGNs), with 50 percent of the galaxy’s flux enclosed within a Petrosian radius $R_{50} \lesssim$~3~arcsec;
(2) high EW(H$\beta$) ($>$~100~\AA) with hot O stars producing ionizing LyC radiation;
(3) high [O~III]$\lambda$4959/H$\beta$ ($\ge$ 1) to ensure galaxies contain high-excitation ionized hydrogen (H\text{\ II}) regions;
(4) sufficiently bright in the far-ultraviolet (FUV) with magnitudes ($\sim$~20.7~mag) in \textit{Galaxy Evolution Explorer} (\textit{GALEX}) and a high enough redshift of $z \sim$~0.3 that allows direct LyC with \textit{HST}-COS; and
(5) high [O~III]$\lambda$5007/[O~II]$\lambda$3727 ratios ($\gtrsim$~5) which may imply density-bounded H\text{\ II} regions [\citenum{2016Natur_529_178I, 2016MNRAS_461_3683I}].
LyC escape was detected in all five galaxies with \textit{HST}-COS.
These properties are often associated with Green Pea-type galaxies.
Since then, Green Pea galaxies have often been found to be LCEs, as shown by several subsequent \textit{HST}-COS studies [\citenum{2018MNRAS_474_4514I, 2018MNRAS_478_4851I, 2021MNRAS_503_1734I}].

\begin{figure}[htbp]
\begin{center}
\begin{tabular}{c}
\includegraphics[width=0.9\textwidth]{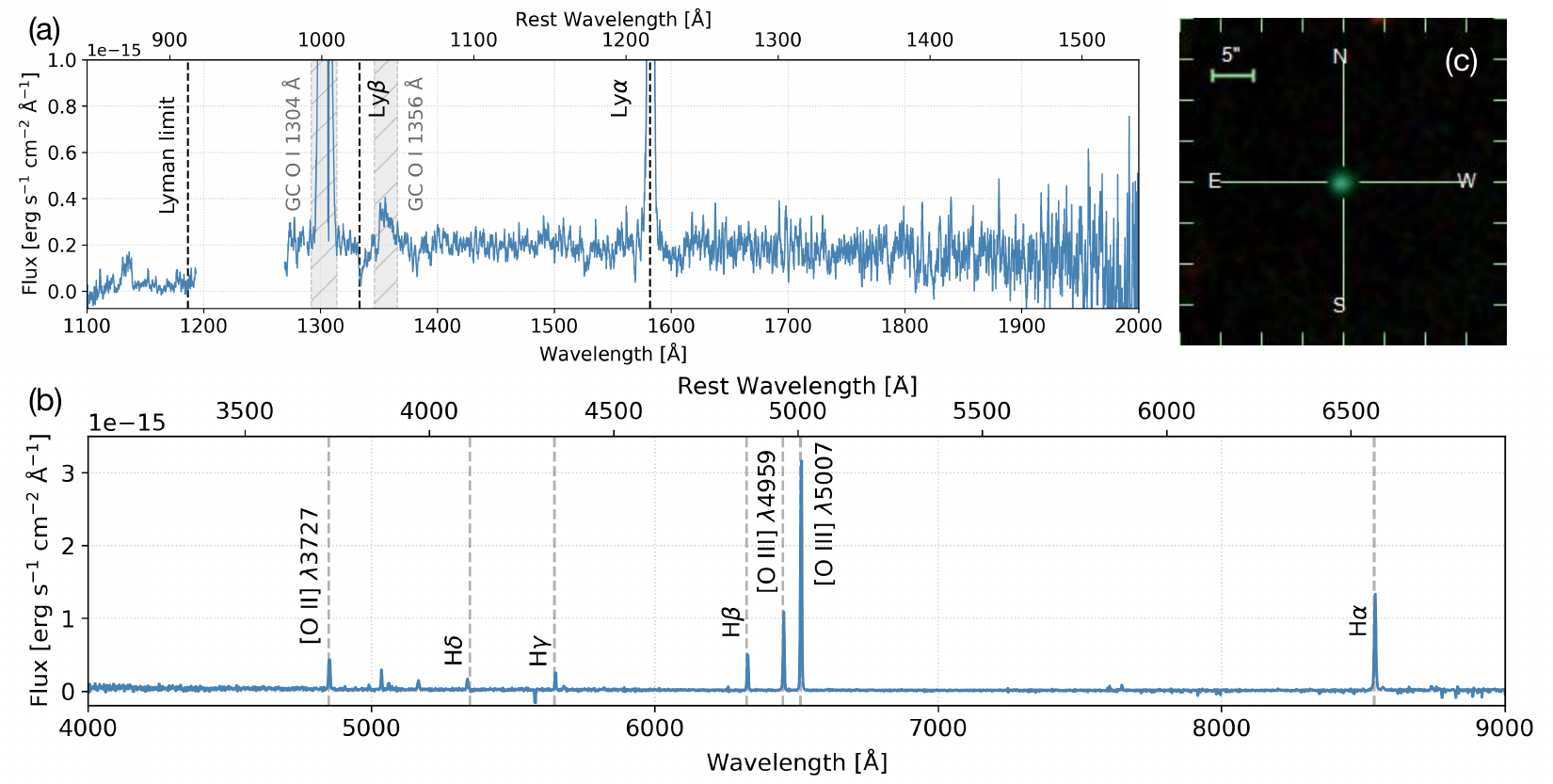}
\end{tabular}
\end{center}


\caption{Example spectra and SDSS image of the Green Pea galaxy J092532+140313 at $z =$ 0.301. (a) \textit{HST}-COS FUV spectrum (program ID: 13744) covering the LyC region with reported leakage [\citenum{2016Natur_529_178I}]. (b) SDSS optical spectrum (ObjID: 1237662301903192106). (c) SDSS $gri$ composite image. Green Pea galaxies appear green in SDSS images due to strong [O~III] $\lambda$5007 emission redshifted into the $r$ band (image credit: SDSS).
\label{fig:GreenPea_Orlitova}}
\end{figure}

The high success rate in detecting LyC leakage in Green Pea galaxies encourages further studies of these and even lower redshift counterparts (e.g., ``blueberry galaxies"; \citenum{2017ApJ_847_38Y}).
Green Pea galaxies are known for their compact size and green color in the SDSS images (Fig.~\ref{fig:GreenPea_Orlitova}).
These objects appear green in SDSS images because the strong [O~III] $\lambda$5007 emission line is redshifted into the SDSS $r$ band, which appears green in color composites.
High [O~III]/[O~II] is indicative of a highly ionized interstellar medium (ISM), with intense ionizing radiation.
The reliability of this metric as an indicator of LyC escape has made it a strong search criterion [\citenum{2013ApJ_766_91J,2014MNRAS_442_900N}].

The Low-redshift Lyman Continuum Survey (LzLCS; \citenum{2022ApJS_260_1F, 2022ApJ_930_126F}) is the culmination of these successes --- 134 orbits of \textit{HST}-COS targeting a variety of metrics for predicting LyC Escape.
They include high [O~III]/[O~II] ratios, high star formation rate surface density $\Sigma_{\rm{SFR}}$, highly negative UV continuum slope $\beta$, and other indicators of strong ionizing potential, including the selection criteria adopted in Refs.~\citenum{2016Natur_529_178I} and \citenum{2016MNRAS_461_3683I}.
The LzLCS studied a sample of 66 nearby star-forming galaxies at $z \sim$ 0.3 as Lyman continuum emitter (LCE) candidates,
with the redshift selected so as to shift the LyC into the \textit{HST}-COS sensitivity range.
Using \textit{HST}-COS with the G140L grating in the \texttt{CENWAV=800} mode [\citenum{2016PASP_128j5006R}], the LzLCS science team analyzed the FUV spectra of the candidates and identified 35 LCEs out of the 66 targets.
This survey represents the largest sample of low-redshift LyC sources to date and provides an unprecedented opportunity for detailed studies of the galaxies and their properties.

\subsection{The Lyman Continuum Escape Fraction} \label{subsec:LyCesc}

\begin{figure}[htbp]
\begin{center}
\begin{tabular}{c}
\includegraphics[width=0.95\textwidth]{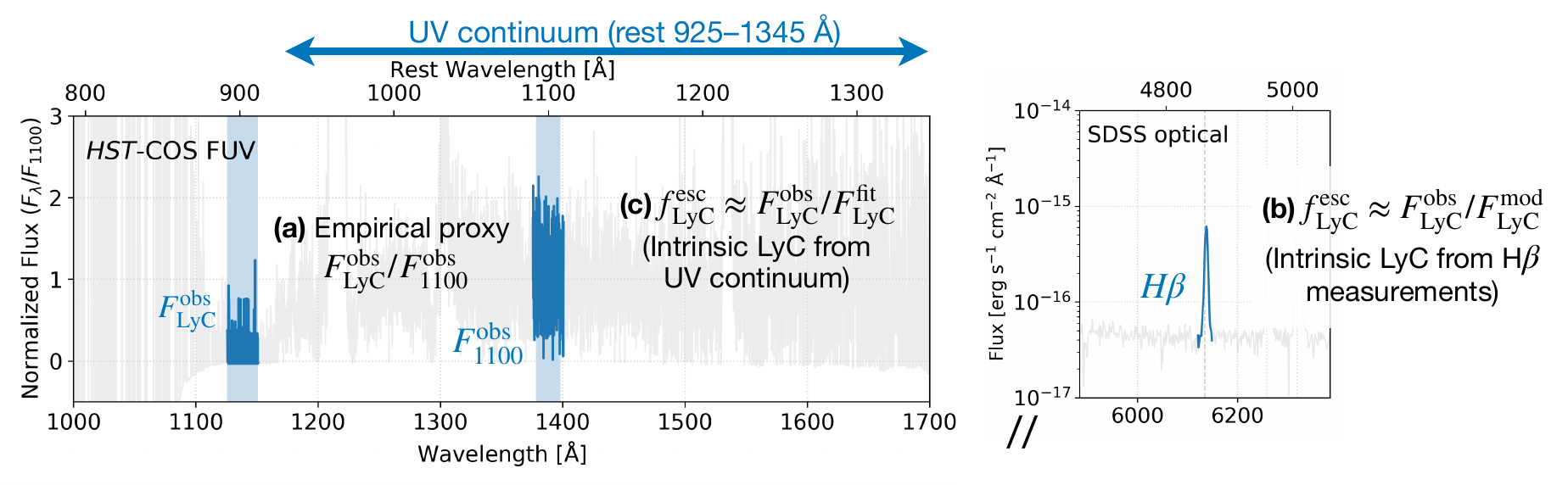}
\end{tabular}
\end{center}


\caption{Illustration of the three methods used to measure the LyC escape fraction in the LzLCS sample (see \S~\ref{subsec:LyCesc}). (a) Empirical proxy using the direct observed flux ratio (Method A). (b) Intrinsic LyC estimated from H$\beta$ via SED modeling (Method B). (c) Intrinsic LyC estimated from fitting the UV continuum (Method C). The \textit{HST}-COS FUV spectra are shown on the left and the SDSS optical spectra on the right.
\label{fig:LzLCS_methods}}
\end{figure}

The LyC escape fraction of galaxies is theoretically defined as:
\begin{equation}
f^{\rm{esc}}_{\rm{LyC}} = \frac{\int^{912}_0 F_{\lambda}^{\rm{obs}}/E_{\lambda} \,d\lambda}{\int^{912}_0 F_{\lambda}^{\rm{int}}/E_{\lambda} \,d\lambda},
\end{equation}
where $F_{\lambda}^{\rm obs}/E_{\lambda}$ represents the observed escaping photon flux, and $F_{\lambda}^{\rm int}/E_{\lambda}$ is the intrinsic photon flux derived from using the spectral energy distribution (SED; equation (3) in Ref.~\citenum{2017ApJ_845_111M}).
Typically, LyC measurements are most sensitive to wavelengths down to $\sim$~700~\AA, where most of the ionizing photons produced by stellar populations are emitted.
Moreover, intrinsic LyC flux from SED also differs from model to model.
Therefore, the LzLCS [\citenum{2022ApJS_260_1F}] summarized and adopted three common methods for obtaining $f^{\rm{esc}}_{\rm{LyC}}$ (Fig.~\ref{fig:LzLCS_methods}):

Method A: Measuring the direct flux ratio of $F^{\rm{obs}}_{\rm{LyC}}/F^{\rm{obs}}_{\rm{1100}}$ as an empirical proxy for $f^{\rm{esc}}_{\rm{LyC}}$, using 1100 \AA\ as the non-ionizing flux [\citenum{2001ApJ_546_665S, 2019ApJ_885_57W}];
 
Method B: Estimating $F^{\rm{int}}_{\rm{LyC}}$ from SED fitting using the observed H$\beta$ flux ($F^{\rm{mod}}_{\rm{LyC}}$), such that $f^{\rm{esc}}_{\rm{LyC}} (\rm{H\beta})$ = $F^{\rm{obs}}_{\rm{LyC}}/F^{\rm{mod}}_{\rm{LyC}}$     [\citenum{2016MNRAS_461_3683I,2018MNRAS_474_4514I}]; and

Method C: Fitting SED models to the UV continuum and computing $f^{\rm{esc}}_{\rm{LyC}}$(UV) = $F^{\rm{obs}}_{\rm{LyC}}/F^{\rm{fit}}_{\rm{LyC}}$ [\citenum{2019ApJ_882_182C}].

The LzLCS follow-up studies further constrained $f^{\rm{esc}}_{\rm{LyC}}$ with other measured quantities from the galaxy spectra.
An analysis of H\text{\ I} (Lyman series) and metallic low ionization state lines estimated the escape fraction of ionizing photons [\citenum{2022AA_663A_59S}].
Observations of FUV stellar continuum slope at 1550~\AA\ ($\beta_{\rm{obs}}^{1550}$) were used to scale the flux ratio of [O~III]/[O~II] and $f^{\rm{esc}}_{\rm{LyC}}$, which enables the evaluation of ionizing photon emissivity at high redshift $z \sim$~4--8 [\citenum{2022MNRAS_517_5104C}].

The results of the LzLCS survey have identified additional unexplored parameter space for LyC studies that could further advance our understanding of these potential proxies for the EoR.
Unfortunately, the limits of \textit{HST} make a follow-up study challenging, both because of the observatory's highly competitive time allocation and to instrumental sensitivity that declines rapidly below 1100~\AA\ [\citenum{2010ApJ_709L_183M}], restricting observations at even lower redshifts.

The Supernova remnants and Proxies for Re-Ionization Testbed Experiment (SPRITE) SmallSat will extend LyC studies into the FUV spectrum and map FUV emission from supernova remnants (SNRs) in the Milky Way, the Magellanic Clouds, and nearby galaxies [\citenum{2019SPIE11118E_0UF, 2022NatAs_6_1213F, 2023SPIE12678E_06I, 2024SPIE1309332}].
In this paper, we analyze the predicted performance for the SPRITE Ionizing
Radiation Emitter Survey (SPIRES) using several known LyC leakers to demonstrate that the mission's pre-launch calibration meets the science objectives. 
This work establishes the commissioning plan ahead of launch in October 2026, and lays the groundwork for the mission science operations phase of the SPIRES survey.


Section \ref{sec:overview_SPRITE} provides an overview of the upcoming SPRITE mission and its science goals.
Section \ref{sec:SPRITEperformance} describes the selection of LCE commissioning targets and presents the predicted on-orbit performance of SPRITE, including its effective area, scattering, in-flight background, and sensitivity in measuring LyC.
Section \ref{sec:commission_plan} outlines the commissioning plan and signal-to-noise estimation for these targets.
Section \ref{sec:degrade} evaluates SPRITE's performance margins under degraded parameters.
Finally, section \ref{sec:summary} concludes the study and outlines ongoing work.

\section{Overview of SPRITE} \label{sec:overview_SPRITE}

\begin{figure}[htbp]
\begin{center}
\begin{tabular}{c}
\includegraphics[width=0.9\textwidth]{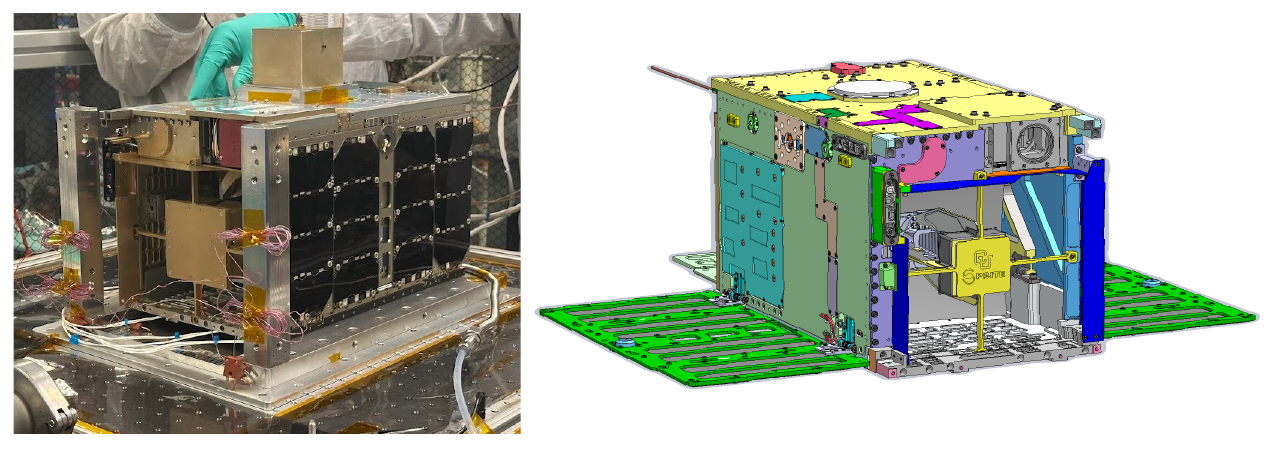}
\end{tabular}
\end{center}


\caption{The SPRITE 12U SmallSat (left) and a CAD model with deployed solar panels and UHF antenna (right).
\label{fig:SPRITE_Bowen}}
\end{figure}

SPRITE is a NASA-funded 12U SmallSat FUV (1000--1750~\AA; Fig.~\ref{fig:SPRITE_Bowen}) imaging spectrograph, developed at the Laboratory for Atmospheric and Space Physics (LASP) at the University of Colorado Boulder [\citenum{2019SPIE11118E_0UF, 2022NatAs_6_1213F, 2023SPIE12678E_06I, 2024SPIE1309332}].
SPRITE has two key science goals:
\begin{enumerate}
    \item To study the mechanisms of LyC escape from low-redshift star-forming galaxies (0.16~$< z <$~0.4) that will help understand how the first galaxies during the EoR ionized the universe (SPIRES); and
    \item To provide detailed spectral mapping of FUV emissions from shocked regions in SNRs in the Milky Way and Magellanic Clouds, as well as from star-forming regions in nearby galaxies ($z<$~0.01; see Ref.~\citenum{carlson2025push}).
\end{enumerate}

\begin{figure}[htbp]
\begin{center}
\begin{tabular}{c}
\includegraphics[width=0.95\textwidth]{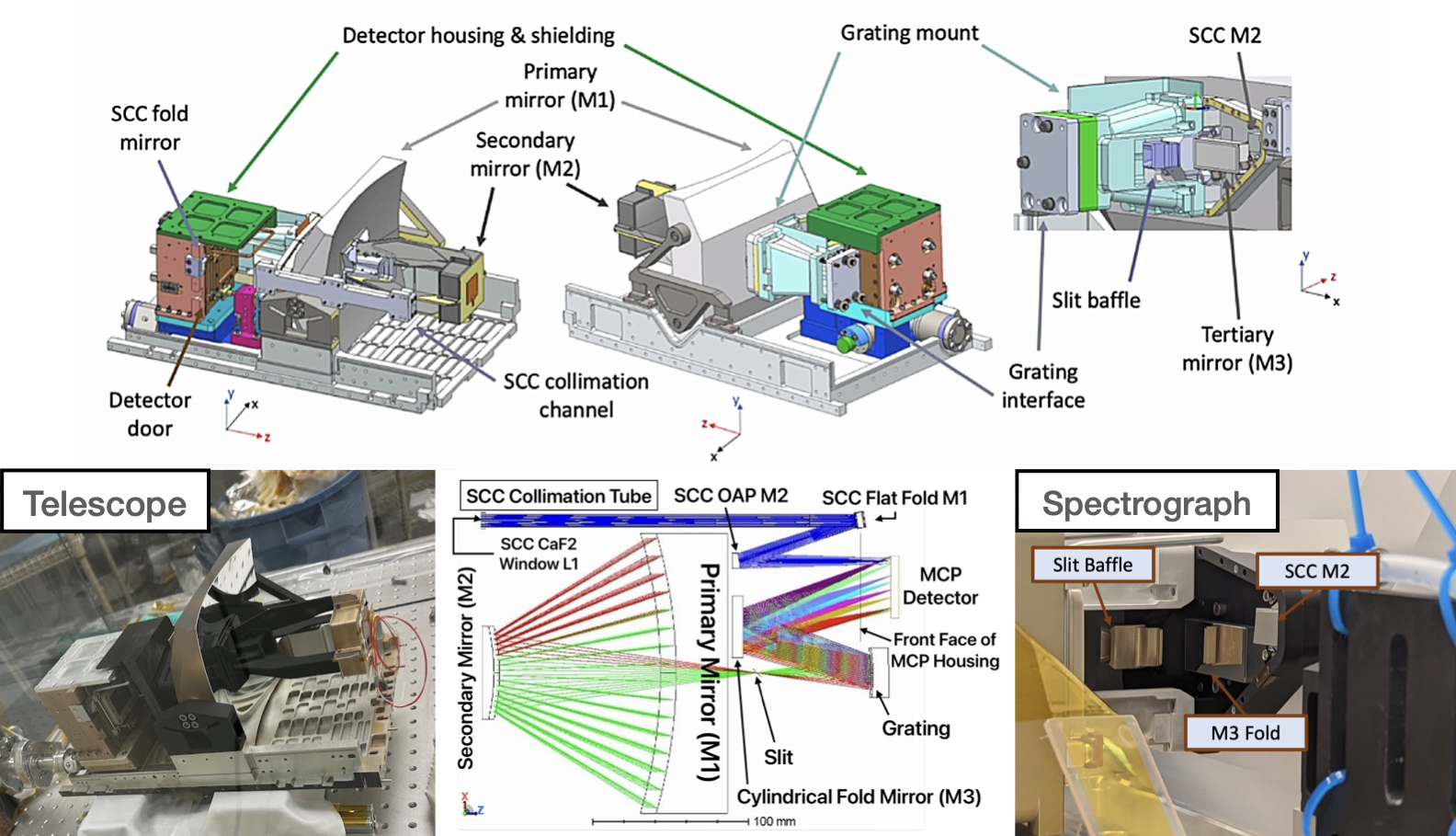}
\end{tabular}
\end{center}

\vspace{5pt} 

\caption{Overview of the SPRITE instrument. Top panels show the front (left) and rear (center) views of SPRITE, as well as the interior view of the spectrograph (right). Bottom panels show the telescope and spectrograph assemblies, along with a ray-trace diagram. Incoming light reflects off M1 onto M2, passes through the slit and grating, and is reflected by M3 onto the MCP detector [\citenum{2023SPIE12678E_0AB}].
\label{fig:SPRITE_model}}
\end{figure}

The SPRITE science instrument comprises a Cassegrain F/2.7 telescope, a grating, a cylindrical fold mirror, and a microchannel plate (MCP) detector encased in a hermetically sealed housing with a one-time-open door.
The instrument is also equipped with the SPRITE calibration channel (SCC), a co-aligned 1350--2000~\AA\ low-resolution imager used to monitor the stability of the detector and mirror coatings in orbit (see Fig.~\ref{fig:centralbulge}).
Fig.~\ref{fig:SPRITE_model} displays a detailed look inside the instrument, with a ray trace diagram in the bottom central panel [\citenum{2023SPIE12678E_0AB}]. A detailed overview of the SPRITE instrument and spacecraft design is presented in prior references [\citenum{2023SPIE12678E_06I,2024SPIE13093E_34B}].

SPRITE is designed as a testbed for new mirror coating methods that are enabling for future Lyman UV ($\lambda <$~1200~\AA) missions, such as the Habitable Worlds Observatory (HWO).
The secondary (M2) and the tertiary fold mirror (M3) are coated with enhanced lithium fluoride protected aluminum (eLiF) coatings that are topped with a thin layer of magnesium fluoride (MgF$_2$) applied with atomic layer deposition (ALD) [\citenum{2017SPIE10401E_19H, 2017ApOpt_56_9941F}].
The primary mirror (M1), with dimensions of 18~$\times$~16~cm$^{2}$, was too large for the eLiF-enabled coating chamber at NASA Goddard Spaceflight Center (GSFC), so it was coated with conventional LiF+Al, but protected with an ALD overcoat for humidity resilience.
The SPRITE grating --- while originally coated with eLiF --- was stripped and replaced with xenon difluoride-catalyzed LiF+Al (XeLiF) in 2024, as XeLiF is the current leading contender coating for HWO  [\citenum{2022SPIE12188E_1VQ}].

\begin{figure}[htbp]
\begin{center}
\begin{tabular}{c}
\includegraphics[width=0.6\textwidth]{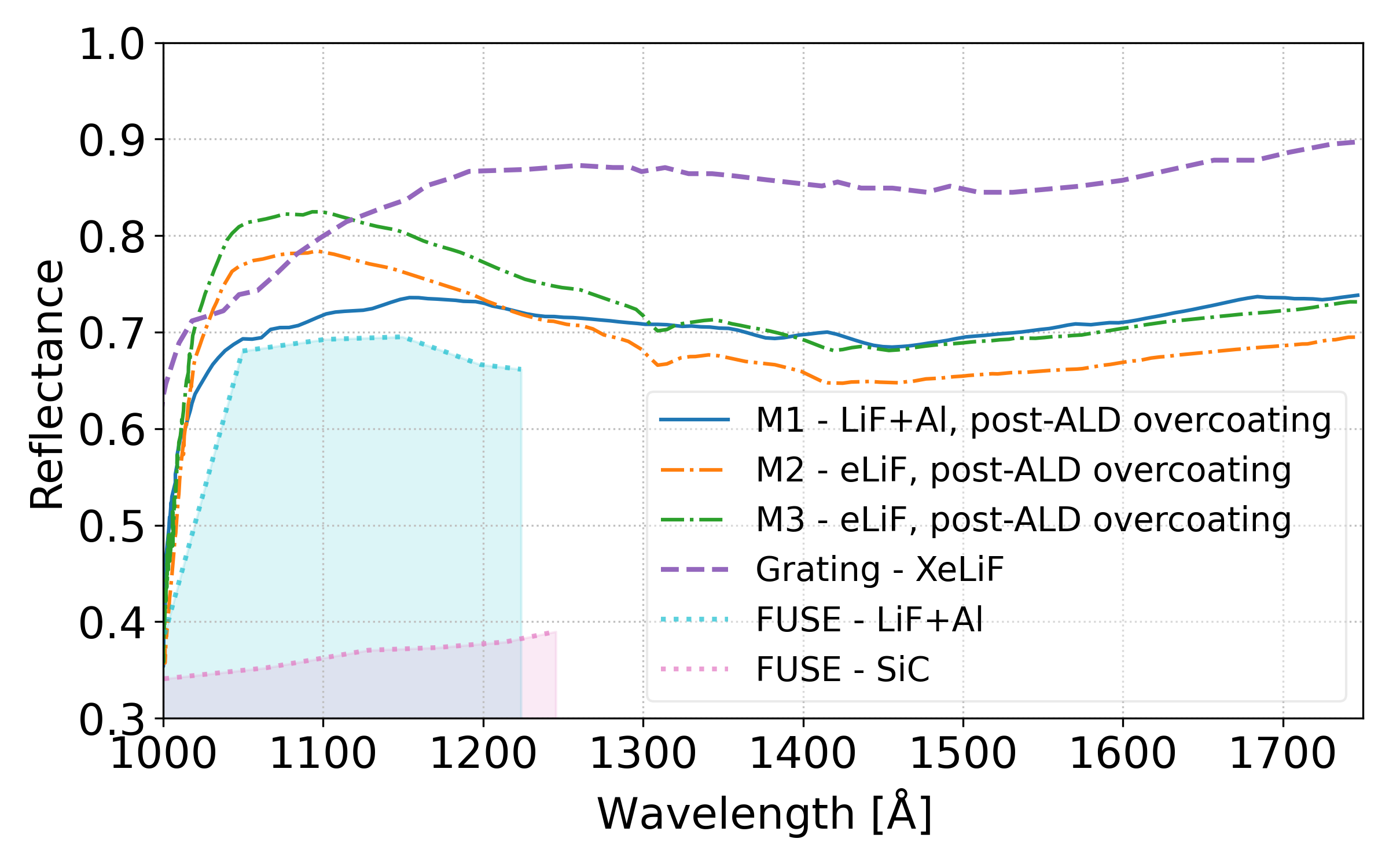}
\end{tabular}
\end{center}


\caption{Reflectance curves of the optical coatings used in SPRITE compared with \textit{FUSE}. The primary mirror on SPRITE uses conventional LiF$+$Al (blue solid line). The secondary and tertiary mirrors use eLiF coatings (orange and green dash-dotted lines). The grating is coated with XeLiF (purple dashed line). The dotted cyan and pink curves show representative \textit{FUSE} LiF$+$Al and SiC coating reflectance, respectively [\citenum{2024SPIE13093E_34B}], with shaded regions to distinguish them from the SPRITE coating curves.
\label{fig:reflectance}}
\end{figure}

The eLiF and XeLiF coatings improve the reflectance at shorter wavelengths from the $\sim$~73\%\ peak achievable with conventional LiF$+$Al (blue solid line in Fig.~\ref{fig:reflectance}) to $\sim$~76--82\%\ (orange and green dash-dotted lines).
The XeLiF method (purple dashed line) offers superior performance in the NUV/optical, which is important for HWO, and appears to be as stable as the protected eLiF enhanced deposition method.
For comparison, \textit{FUSE} used conventional LiF$+$Al coatings and SiC coatings, shown as dotted curves in Fig.~\ref{fig:reflectance}.
The LiF$+$Al coating reached $\sim$~70\%\ reflectance over part of the \textit{FUSE} bandpass, while the SiC coating achieved $\sim$~30--40\%\ reflectance when fresh, but degraded by $\sim$~10\%\ within 9 days of atmospheric exposure [\citenum{1996SPIE_2807_172K}].
\textit{HUT} mirror reflectance is not shown because comparable coating-reflectance data were not available; however, its lower-reflectivity optics are reflected in the effective-area comparison shown in Fig.~\ref{fig:Aeff_SPRITE_Bowen}.
SPRITE's advanced coatings provide higher reflectance over much of the Lyman-UV bandpass.

The low-background, photon-counting MCP detector of SPRITE, with dimensions of 39~$\times$19 mm$^2$ and a plate scale of 144 arcsec mm$^{-1}$, also enhances the sensitivity of SPRITE relative to prior instruments, with an in-lab background of $<$~0.1 counts cm$^{-2}$ s$^{-1}$.
These technologies provide high FUV sensitivity comparable to that of much larger FUV space missions, such as \textit{FUSE}, but in a compact form.

This paper focuses on pre-launch performance analysis of SPRITE’s first science goal --- the detection and characterization of LyC emission from low-redshift galaxies.
An earlier paper evaluated the performance and commissioning plan for SPRITE's second science objective, push-broom mapping of extended objects (Carlson et al. 2025; \citenum{carlson2025push}).
Since that publication, final instrument calibrations have been completed.
This has led to updates in the instrument performance used in this analysis.
Here, we elaborate on SPRITE's ability to detect low-redshift LyC galaxies, including performance under potential instrument degradation.

\section{Predicted Analysis of SPRITE Performance} \label{sec:SPRITEperformance}

Observations of the Lyman continuum radiation are complicated by both the challenges of observing in this region and the intrinsically low brightness of the escaping ionizing photons.
In order to detect this signal from $z \sim$ 0--0.3, SPRITE requires high sensitivity in the 900--1200~\AA\ bandpass.
Unfortunately, there exists no efficient mirror coating for the 900--1000~\AA\ regime [\citenum{1996SPIE_2807_172K, 2014SPIE_9144E_4HM}].
However, the combination of advanced mirror coatings, a low background detector, and the imaging spectrograph design significantly improves SPRITE's sensitivity from 1000 to 1150~\AA\ relative to the prior state-of-the-art [\citenum{2017ApOpt_56_9941F, 2022SPIE12188E_20R}].
In addition, SPRITE is designed as an orbital testbed for these coatings following successful deployment on prior sounding rocket missions [\citenum{2021SPIE11821E_0HC}], and in advance of potential use on the HWO.
While the use of these coatings restricts SPRITE to observing galaxies at $z >$ 0.15 ($\lambda_{\rm{obs}} >$ 1049~\AA) because of the short-wavelength reflectance limits, it still represents a broader LyC bandpass than is accessible with \textit{HST}-COS.

\begin{figure}[htbp]
\begin{center}
\begin{tabular}{c}
\includegraphics[width=0.95\textwidth]{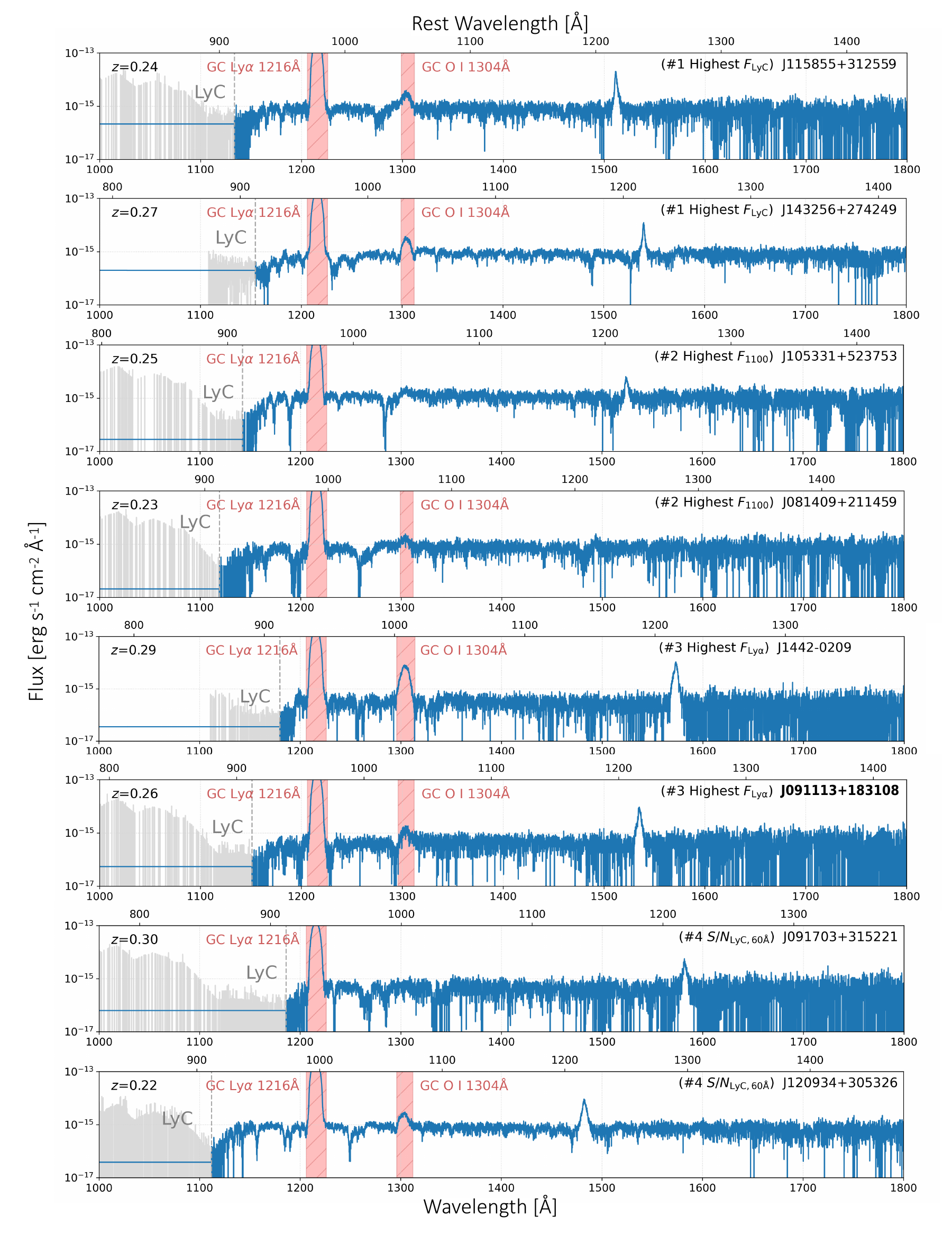}
\end{tabular}
\end{center}


\caption{\textit{HST}-COS G140L spectra of the eight selected LyC commissioning targets, grouped by the selection criteria described in \S~\ref{sec:SPRITEperformance}. The gray lines show the original spectra. The blue lines show LyC-averaged spectra. Red shaded regions indicate geocoronal Lyman-alpha (1216~\AA) and O\text{\ I} 1304~\AA\ emission from the Earth's exosphere and are not associated with the targets. J091113+183108, the target used throughout the paper for the performance analysis, is marked in bold.
\label{fig:8targetHST}}
\end{figure}

To evaluate SPRITE’s predicted performance for LyC observations, we analyze the \textit{HST}-COS spectra of the 35 LCEs identified in the LzLCS survey [\citenum{2022ApJS_260_1F}].
From the list of LCEs, we select eight galaxies to test and demonstrate the capabilities of SPRITE.
Fig.~\ref{fig:8targetHST} showcases these eight commissioning targets selected based on the following criteria:

\# 1: Two targets with the highest LyC flux ($F_{\rm{LyC}}$) as measured in the limited \textit{HST} bandpass (a rest-frame spectral bin of 20~\AA\ as close as possible to $\lambda_{\rm{rest}} =$ 900~\AA; \citenum{2022ApJS_260_1F});

\# 2: Two targets with the highest average continuum flux at rest-frame $\lambda_{\rm{rest}} \sim$ 1100~\AA\ ($F_{1100}$);

\# 3: Two targets with the brightest intrinsic Ly$\alpha$ emission feature ($F_{\rm{Ly\alpha}}$); and

\# 4: Two median LzLCS targets with signal-to-noise ratio $S/N > 4$ in the LyC (60~\AA\ binned).

In Fig.~\ref{fig:8targetHST}, the gray lines show the original \textit{HST}-COS spectra [\citenum{2012ApJ_744_60G}] obtained from the Mikulski Archive for Space Telescopes (MAST), an astronomical data archive for space missions.
The red shaded boxes mark the geocoronal (GC) emission lines --- Lyman-alpha and O\text{\ I} 1304~\AA\ --- that are commonly seen in FUV spectra taken from low earth orbit.
The GC emissions originate primarily from solar radiation resonantly scattering off of hydrogen and oxygen atoms in the Earth's exosphere (or geocorona; \citenum{Mange1972-yl}).
These geocoronal lines are not related to the target galaxies.
We focus on the rest of the spectra outside of the shaded boxes.

The blue lines show the \textit{HST} spectra after masking low signal-to-noise ($S/N$) regions, especially at wavelengths $\lambda <$ 1100~\AA, and measuring the LyC flux in a narrow rest-frame window following the procedure adopted in LzLCS.
\textit{HST}-COS has a rapidly diminishing effective area ($A_{\rm{eff}}$) at wavelengths $\lambda <$ 1150~\AA\ due to the drop-off in reflectance of the MgF2-protected aluminum mirror coatings.
While there remains a significant throughput at shorter wavelengths, it is insufficient to achieve the required $S/N$ in these observations [\citenum{2010ApJ_709L_183M, 2024cosi_book_17H}].
Therefore, we constrain our analysis of the \textit{HST}-COS data to $\lambda >$ 1150~\AA, and average the limited LyC flux available over 1150~\AA\ $\lesssim \lambda < \lambda_{\rm{LyC,obs}}$ to estimate the LyC of the LzLCS galaxies.

We caution that this is a simplified estimate of the LyC flux.
In reality, the LyC flux should not be flat at higher energies beyond the Lyman limit.
The stellar LyC flux may increase due to the decreasing hydrogen ionization cross section (Figure 4 of Ref.~\citenum{2017ApJ_845_111M}), or decrease if there is a narrow bump near 900~\AA\ contributed by free-bound nebular LyC emission, which diminishes at shorter wavelengths [\citenum{2024MNRAS_530_2133S}].
Since which effect dominates is source-dependent and not yet well constrained, we adopt a constant average LyC flux density as a representative approximation for estimating SPRITE’s LyC sensitivity.
These averaged values are comparable to those presented in the LzLCS survey [\citenum{2022ApJS_260_1F}].
Hence, we carry out the analysis of these eight commissioning targets using the LyC-averaged \textit{HST} spectra shown in the blue line in Fig.~\ref{fig:8targetHST}.

To detect a 3$\sigma$ level of LyC from the LCEs, SPRITE requires sensitivity on the order of $\lesssim$ 10$^{-17}$ erg s$^{-1}$ cm$^{-2}$~\AA$^{-1}$ in $F_{\rm{LyC}}$.
Here, this value is not tied to a specific LyC bin size; the predicted $S/N$ for the binned LyC measurements is calculated in Section~\ref{sec:commission_plan}.
The ultimate limiting sensitivity of SPRITE depends on the effective area, background noise, and exposure time.

\subsection{Effective Area of SPRITE} \label{subsec:SPRITE_Aeff}

\begin{figure}[htbp]
\begin{center}
\begin{tabular}{c}
\includegraphics[width=0.65\textwidth]{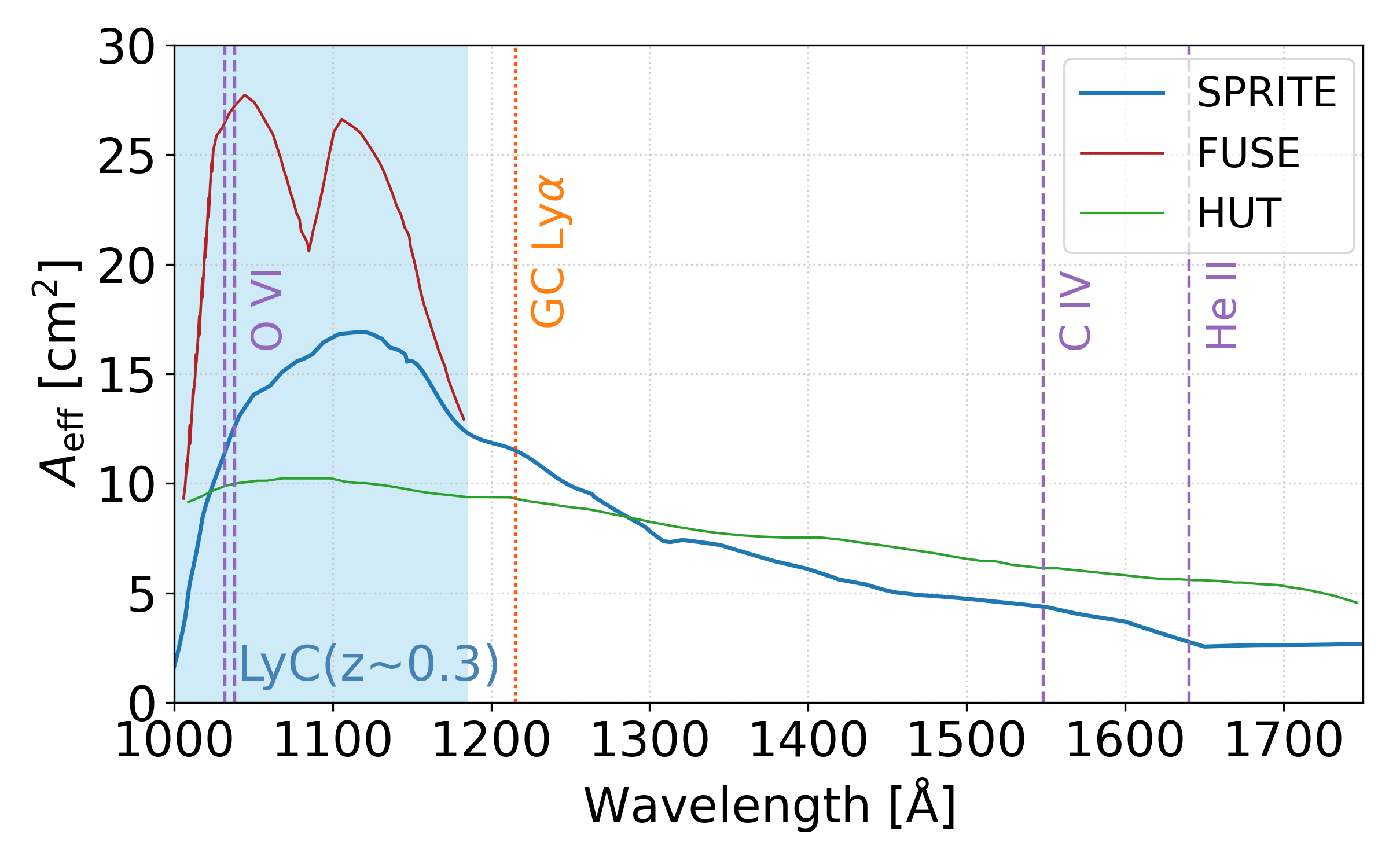}
\end{tabular}
\end{center}


\caption{Effective light-collecting area of SPRITE (blue) compared with \textit{FUSE} (red) and \textit{HUT} (green). The shaded region indicates the LyC wavelength range accessible to SPRITE for galaxies at $z \sim$ 0.3. Dashed purple lines mark common SNR emission lines (O\text{\ VI} $\lambda \lambda$1032, 1038, C\text{\ IV} $\lambda \lambda$1548, 1551, and He\text{\ II} $\lambda$1640). The dotted orange line marks geocoronal Lyman-alpha line.
\label{fig:Aeff_SPRITE_Bowen}}
\end{figure}

The eight commissioning targets span a redshift range of 0.22 $<z<$ 0.30.
In Fig.~\ref{fig:Aeff_SPRITE_Bowen}, we explore the $A_{\rm{eff}}$ of SPRITE compared to that of two major UV telescopes in space --- \textit{HUT} and \textit{FUSE}, both decommissioned in the 1990s and 2000s, respectively.
While \textit{FUSE} has a larger collecting area per channel --- approximately five times that of SPRITE --- it does not extend redward into the FUV beyond 1150~\AA\ nor does it exceed the $A_{\rm{eff}}$ of SPRITE by $\gtrsim$ 2 times over most of the bandpass.
\textit{HUT} has a comparable $A_{\rm{eff}}$ to SPRITE in measuring $F_{1100}$ of these galaxies.
However, SPRITE has an $A_{\rm{eff}}$ typically about 1.5 to 2 times that of \textit{HUT} over the wavelength range where we can observe the LyC spectrum of galaxies in this redshift range.
SPRITE, therefore, is able to achieve an $A_{\rm{eff}}$ in the Lyman UV comparable to or greater than prior larger Explorer-class observatories.
The $A_{\rm{eff}}$ of \textit{HST}-COS is far greater than SPRITE at $\lambda_{\rm{obs}} >$ 1100~\AA.
However, due to the advanced mirror coating (eLiF and XeLiF) on SPRITE, it achieves an $A_{\rm{eff}}$ comparable to the G140L from 1020 to 1080~\AA\ despite less than 1\%\ of the telescope collecting area.

\subsection{Scattered Light and the Predicted In-flight Background} \label{subsec:SPRITE_gratingscatter}
  
A significant factor governing SPRITE’s sensitivity is the instrument background count rate, which is dominated in SPRITE by scattered light within the optical path.
While SPRITE employs an exceptionally low noise detector, laboratory testing revealed that the instrument exhibits scattered light that is directly proportional to the incident FUV light entering the spectrograph.
Many attempts were made to baffle this light, from blocking whole sections of the telescope to inserting light-absorbing materials around the optical assemblies, but with no success. From this, we conclude that the scattered light is primarily coming directly from the optics themselves.
The measured level of average scattered light across the detector is approximately 7~$\times$~10$^{-7}$ of the total intensity per \AA\ --- meaning that for a spectrum with 10 million counts recorded in an integration, we would expect roughly 7 events of scattered light per \AA -sized spectral bin (roughly 45~$\times$~100~$\mu$m).

Upon further investigation using a Hamamatsu H2D2 deuterium lamp (1150--4000~\AA), we find that this scatter arises primarily from two components:
(1) scatter from the SPRITE optics, which are of good quality but manufactured with a relatively modest polishing requirement ($\leq$ 10~\AA\ micro-roughness), and
(2) scatter from the grating blaze facets.
SPRITE is exceptionally sensitive to both effects due to the extremely compact nature of the instrument.
As light scatters in a cone-shaped pattern from each optic, the scattered light spreads $\propto$~1/$r^{2}$, where $r$ is the distance from the scattering optic.
Therefore, a compact instrument with a given detector size will record far more background-scatter events than a longer-focal-length instrument with the same detector area. As an example, an instrument with a 25~cm focusing distance from the last optic would have $\approx$~9\%\ of the scattered light as SPRITE with its $\approx$~7.5~cm separation between M3 and the detector.

\begin{figure}[htbp]
\begin{center}
\begin{tabular}{c}
\includegraphics[width=0.7\textwidth]{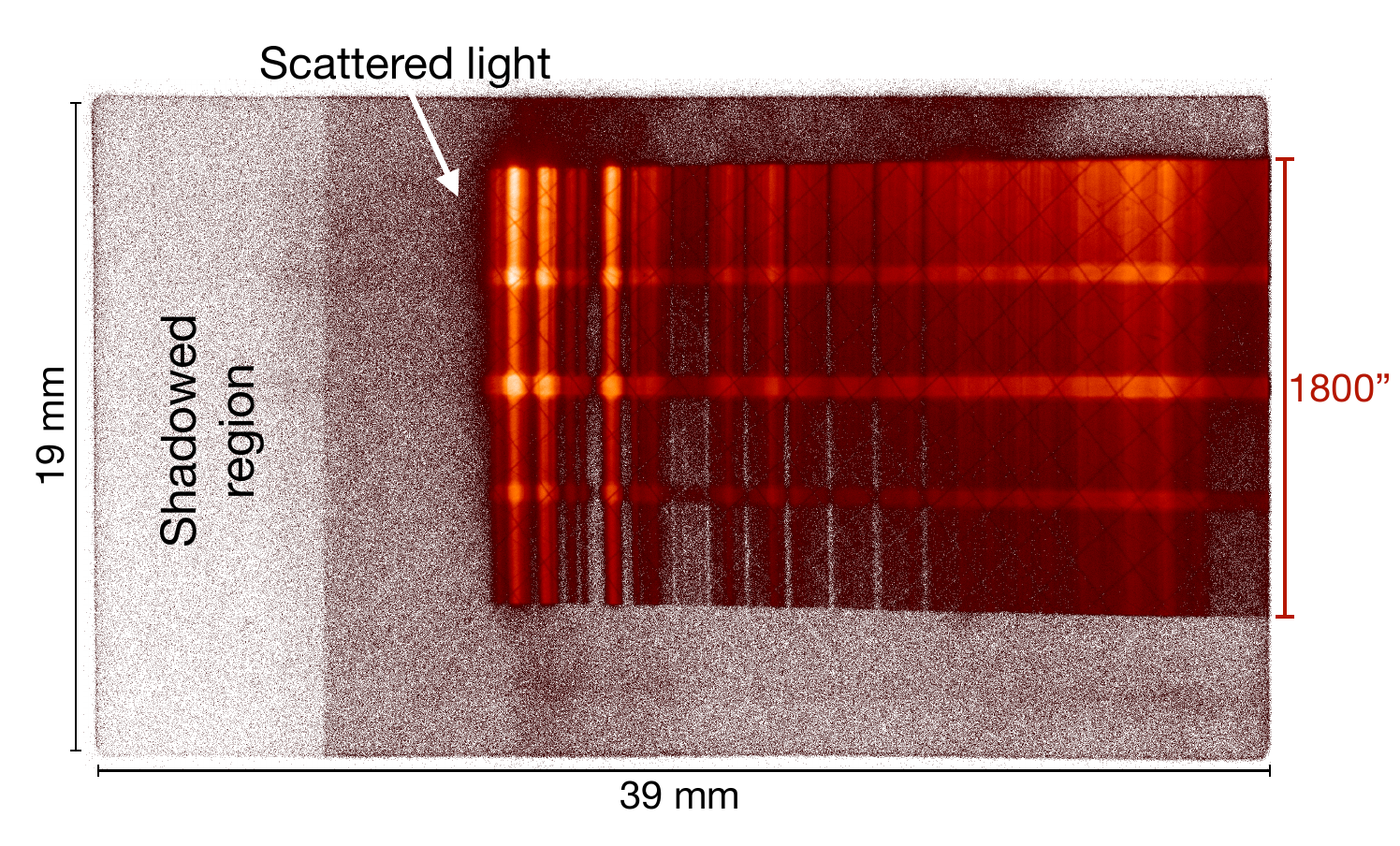}
\end{tabular}
\end{center}


\caption{
SPRITE detector image showing the dispersed spectral light from a deuterium lamp (bright red) and surrounding scattered light (gray). The left portion of the detector lies in the shadowed region with no direct line of sight to the grating; any signal detected there arises solely from scattered light from M3. The area between the shadowed region and the bright spectrogram corresponds to approximately 1000--1150~\AA, up to the MgF$_2$ window cutoff of the D$_2$ lamp. The full detector area is 39~$\times$~19~mm$^2$.
\label{fig:scatter}}
\end{figure}

Most of this scatter appears to originate from the grating rather than M3 or the telescope. A portion of the SPRITE detector is shadowed by baffles and has a line of sight to M3, but not to the grating.
To reach this shadowed detector region, light must scatter directly from M3 due to the position of baffles and other structures (Fig.~\ref{fig:scatter}). This region therefore provides an estimate of the scatter due to M3, with the remainder we attribute to the grating (though the telescope also likely contributes, the slit limits that impact).

The remaining detector area has a background level roughly 7 times that within the shadowed region.
If the power is attributed entirely to the grating, we find that it has a very similar scatter performance to the blazed holographic gratings on \textit{HST}-COS, which scattered $\sim$ 1.6~$\times$~10$^{-5}$ of the on-axis incident intensity (I/I$_{0}$) at a physical distance of approximately 6 mm on the detector [\citenum{2000SPIE_4013_360O}].
The method of measurement differs between SPRITE and COS; however, the values are similar. This effect was less apparent on COS because, as a point-source spectrograph, it only subtends a small part of the sky, limiting the amount of bright geocoronal Lyman-alpha that enters the spectrograph.
While SPRITE has $\sim$~1\%\ of the collecting area of \textit{HST}, SPRITE's long slit covers 3700 times the area of the sky, yielding roughly 37 times the geocoronal Lyman-alpha into the system.

The combination of what appears to be a normal level of holographic blazed grating scatter, along with the compact geometry and large field of view, makes this scattered Lyman alpha the dominant background source on SPRITE.
For a nighttime observation with a geocoronal brightness of $\sim$~2 kR, where 1 R = 10$^{6}$/4$\pi$ photons cm$^{-2}$ s$^{-1}$ sr$^{-1}$, depending on orbit and solar cycle, and using SPRITE's 1800$^{\prime\prime} \times$10$^{\prime\prime}$ slit, we expect roughly 1300 events s$^{-1}$ of Lyman alpha on the SPRITE detector.
At the scatter profiles measured, this amounts to a predicted equivalent background of 3 to 9 counts cm$^{-2}$ s$^{-1}$ across the detector, up from a formulation phase prediction of 0.5.
This background has little impact on the SPRITE galaxy or supernova remnant science, but will dominate the ultimate sensitivity of the instrument for Lyman continuum measurements.

\subsection{Sensitivity of SPRITE} \label{subsec:SPRITE_sensitivity}

\subsubsection{Detector Dark Rate and Noise Budget} \label{subsubsec:detector_darkrate}

With the instrument background characterized in Section~\ref{subsec:SPRITE_gratingscatter}, we now estimate SPRITE’s achievable sensitivity for LyC observations.
SPRITE’s imaging-spectrograph design yields a tight point-source spectrum with a small extraction area on the detector, thereby minimizing background.
Taking into account the nominal background and effective area, we compute the expected signal-to-noise ratio and the resulting LyC sensitivity for the LCE targets.


The laboratory performance of the SPRITE MCP detector exhibits $<$~0.1 counts cm$^{-2}$ s$^{-1}$ in a dark image, which translates to $\sim$~10$^{-7}$ counts~pixel$^{-1}$ s$^{-1}$, or roughly one event per 11 $\times$ 14 $\mu$m$^{2}$ digital ``pixel" per 100 days.
Once on-orbit, this intrinsic dark rate will be surpassed by secondary X-rays produced by $\beta$ radiation being absorbed by the SPRITE structures, raising the predicted rate to approximately 0.5--1.0 counts cm$^{-2}$ s$^{-1}$, or $\sim$~1 event per pixel per 2 weeks.
This rate has been shown to be proportional to the mass of the observatory [\citenum{2020SPIE11454E_1HS}], for which SPRITE's small mass may be near-negligible. Unfortunately, laboratory measurements of the instrument scatter indicate that in the compact geometry of SPRITE with the wide field of view, scattered Lyman alpha will dominate over these factors, raising the final rate to between 3 and 8 counts~cm$^{-2}$ s$^{-1}$, depending on the orbital phase and other factors (see \S~\ref{subsec:SPRITE_gratingscatter}).
Given this, we adopt a nominal dark rate of $D_0 =$ 4 counts cm$^{-2}$ s$^{-1}$ for SPRITE's nominal case, or roughly an event per pixel every 40 hours.
The SPRITE detector has no read noise; therefore, this and shot noise are the only noise terms considered in the final analysis.

\subsubsection{Line-Spread Function and Extract Aperture} \label{subsubsec:linespread}

\begin{figure}[htbp]
\begin{center}
\begin{tabular}{c}
\includegraphics[width=0.65\textwidth]{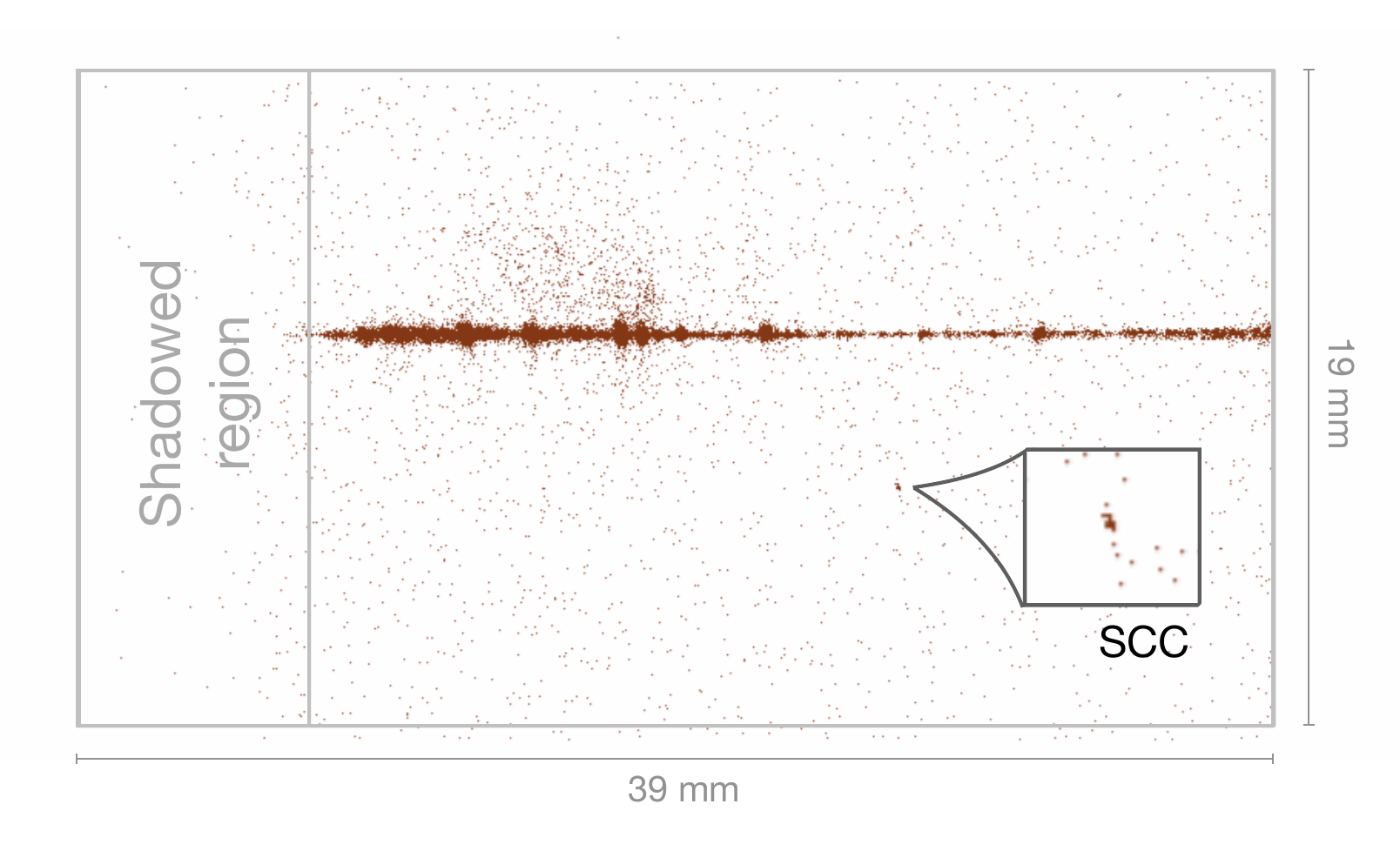}
\end{tabular}
\end{center}

\vspace{-5pt} 

\caption{Spectrum of an air-fed plasma arc lamp measured at the central bulge of SPRITE’s long slit. The circle indicates the position of the SCC on the detector.
\label{fig:centralbulge}}
\end{figure}

\begin{figure}[htbp]
\begin{center}
\begin{tabular}{c}
\includegraphics[width=0.9\textwidth]{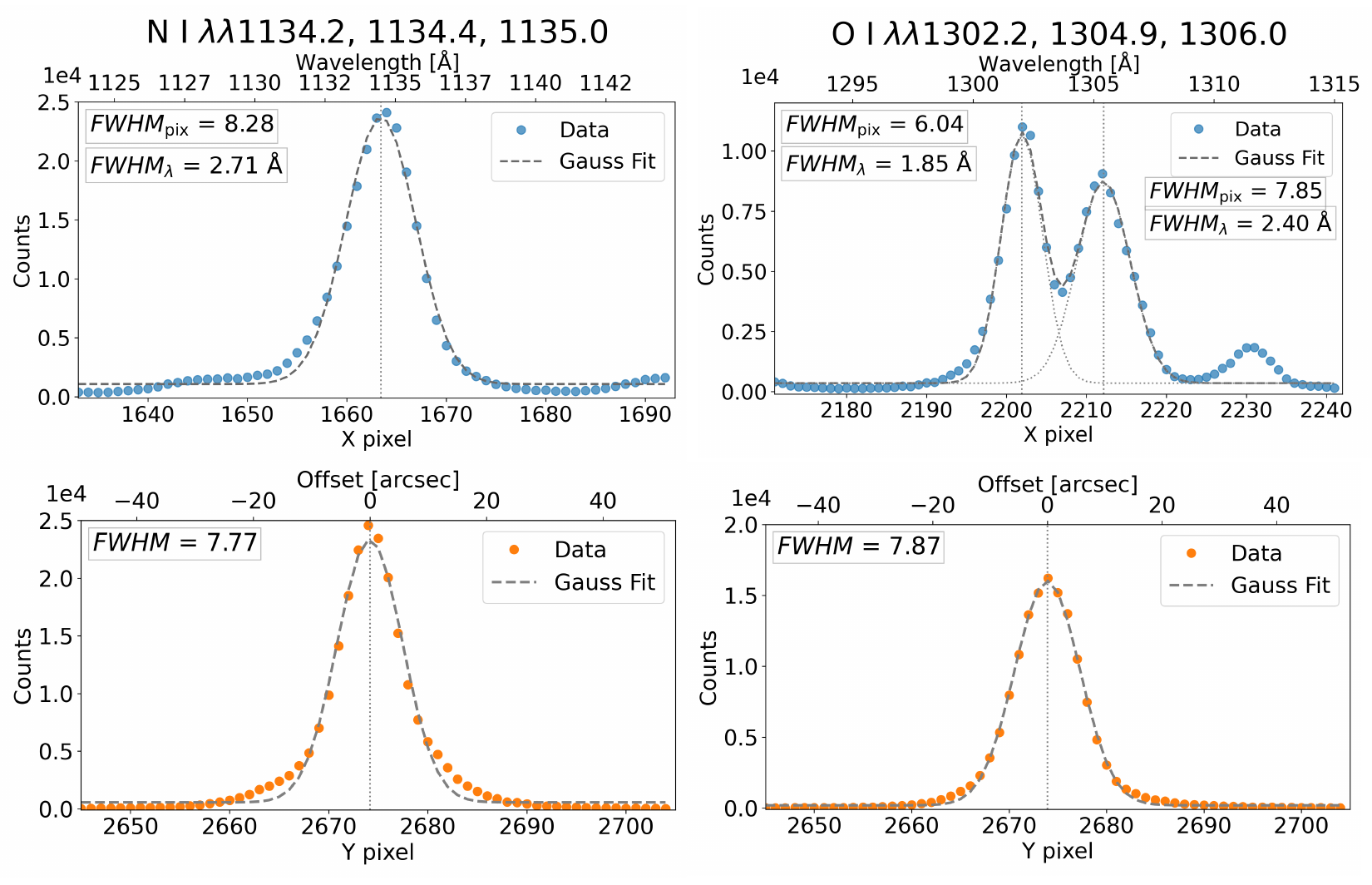}
\end{tabular}
\end{center}


\caption{SPRITE laboratory line-spread profiles in the spectral (top) and spatial (bottom) directions.
Left: N I $\lambda\lambda$1134.2, 1134.4, 1135.0, corresponding to observed LyC region wavelengths for SPRITE targets at $z >$ 0.2.
Right: O~I $\lambda$1302.2 and O~I $\lambda\lambda$1304.9, 1306.0, demonstrating multi-component resolution wavelengths corresponding to observed $F_{1100}$.
The measured spectral FWHM is $\sim$~2.7~\AA\ near the LyC region and $\sim$~1.8 to 2.3~\AA\ near the observed $F_{1100}$; the spatial FWHM is $\sim$~8 pixels.
The feature near $x \sim $ 2230 pixels is not associated with the O~I emission lines and is not included in the analysis.
\label{fig:FitLineExample}}
\end{figure}

We next characterize the spectral and cross-dispersion widths that set the signal and background extraction apertures.
Fig.~\ref{fig:centralbulge} shows the spectrum of an air-fed plasma arc lamp illuminating the central bulge of SPRITE's long slit, based on CU/LASP far-ultraviolet calibration facilities and procedures described in Ref.~\citenum{2016JAI_540001F}.
The SCC location is marked for reference.
We use these laboratory data to measure the line-spread profile in both the spectral and spatial direction using the laboratory measurements shown in Fig.~\ref{fig:FitLineExample}.

We focus on the emission lines of N~I $\lambda\lambda$1134.2, 1134.4, 1135.0, which fall near the observed LyC region wavelengths for SPRITE's targets at $z \sim 0.3$.
We also present the emission lines of O~I $\lambda\lambda$1302.2, 1304.9, 1306.0 to demonstrate SPRITE's capability to resolve two components --- O~I $\lambda$1302.2 and O~I $\lambda\lambda$1304.9, 1306.0.

Our measurements indicate a spectral FWHM of approximately 2.71~$\pm$~0.05~\AA\ in the observed LyC region wavelengths and $\sim$~1.85~$\pm$~0.04 to 2.40~$\pm$~0.06~\AA\ at wavelengths corresponding to observed $F_{1100}$ region at the redshifts of these targets.
The cross-dispersion FWHM is $\sim$~7.8~$\pm$~0.1 pixels, corresponding to $\sim$~13~$\pm$~0.2 arcsec.
These measured values are consistent with SPRITE's 4~\AA\ spectral-resolution requirement and are close to the flight objective of $<$~2~\AA\ at the longer $F_{1100}$ wavelengths.
The 2.7~\AA\ FWHM measured near the observed LyC region is broader than the idealized raytrace value, but still satisfies the mission requirement [\citenum{2023SPIE12678E_0AB}].
These measured values are partially limited by the resolution limit of $\sim$ 4$^{\prime\prime}$ of the vacuum collimator generating these signals in the laboratory.




\begin{figure}[htbp]
\begin{center}
\begin{tabular}{c}
\includegraphics[width=0.65\textwidth]{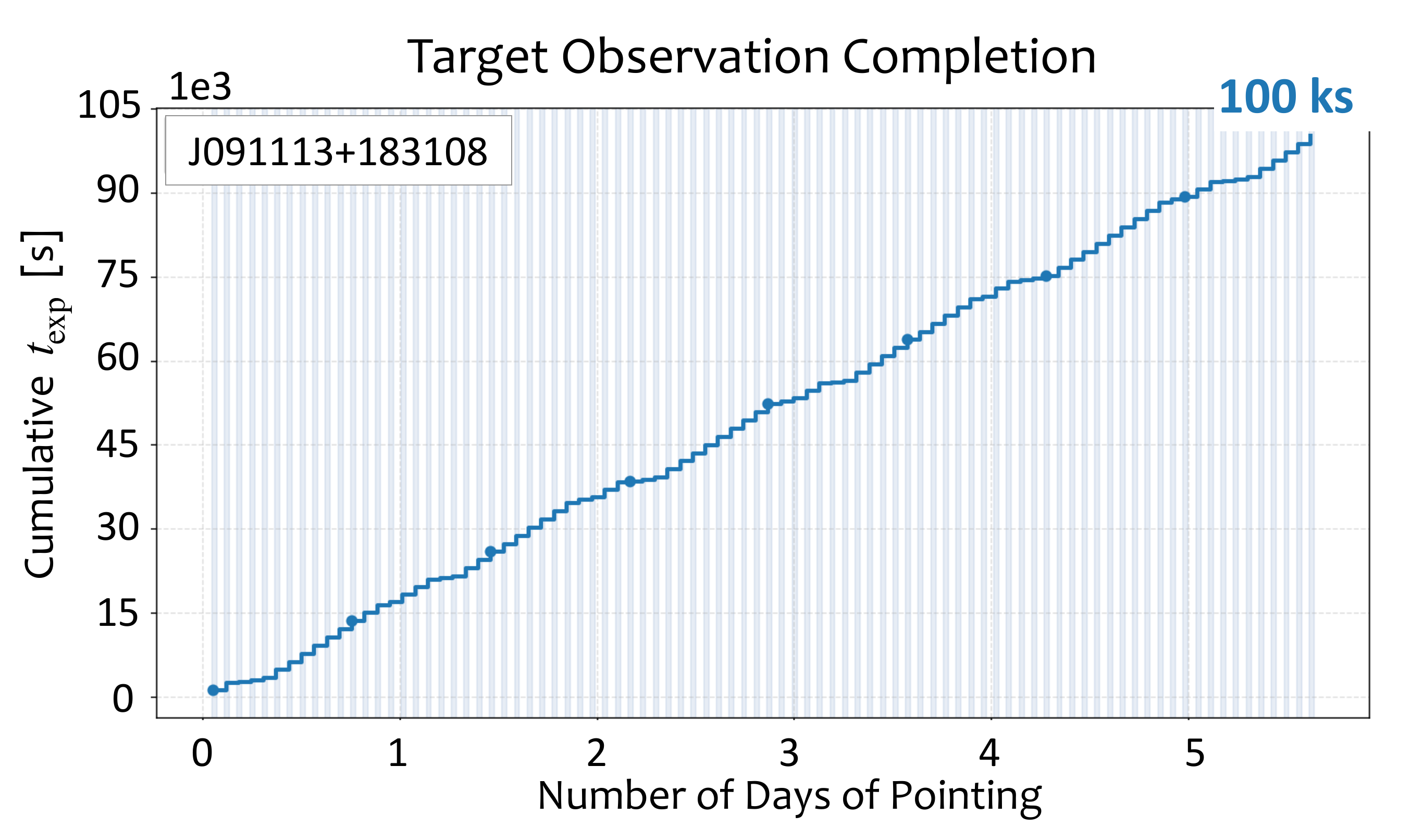}
\end{tabular}
\end{center}


\caption{A simulation of mission planning for accumulating 10$^5$ s of exposure time on target J091113$+$183108. Gray and white vertical bands represent individual orbit. In this case, J091113$+$183108 requires $\sim$~90 orbits to observe for 10$^5$ s, with some portion of each orbit used for a secondary target when J091113$+$183108 is not available.
\label{fig:SPRITE_1target_TotalTime}}
\end{figure}

The $F_{\rm{LyC}}$ for our commissioning targets (Fig.~\ref{fig:8targetHST}) ranges between 10$^{-17}$ and 10$^{-16}$ erg s$^{-1}$ cm$^{-2}$~\AA$^{-1}$.
Based on these fluxes, we assume an exposure time of $t_{\rm{exp}} =$ 10$^5$ s (27.78 hours) for the purpose of modeling the SPRITE observations, though in reality each target may have a dynamic exposure time.
As the observations will be spread over several days (Fig.~\ref{fig:SPRITE_1target_TotalTime}), with partial data downlinks in between, we can choose to extend or truncate observations as the data merit.
This will be challenging, however, as the data completeness factor for a given S-band downlink will not be 100\%, often requiring several passes to fill in dropped packets and reach the $>$ 95\%\ completeness level for a single observation [\citenum{2023AJ_165_64E, 2023AJ_165_63F}].
With observations spread over multiple days, and each day of observations itself taking multiple days to reach 100\%\ completeness, we will not be able to effectively analyze the signal-to-noise in the LyC until likely most or all of the 10$^5$ s exposure time has passed. 
We expect this to improve as the SPRITE downlink, commissioning, and data pipeline are refined over the mission, but for the purposes of commissioning and this paper, we set a fixed exposure time, with the option to add exposure time after the initial analysis.

\begin{figure}[htbp]
\begin{center}
\begin{tabular}{c}
\includegraphics[width=0.55\textwidth]{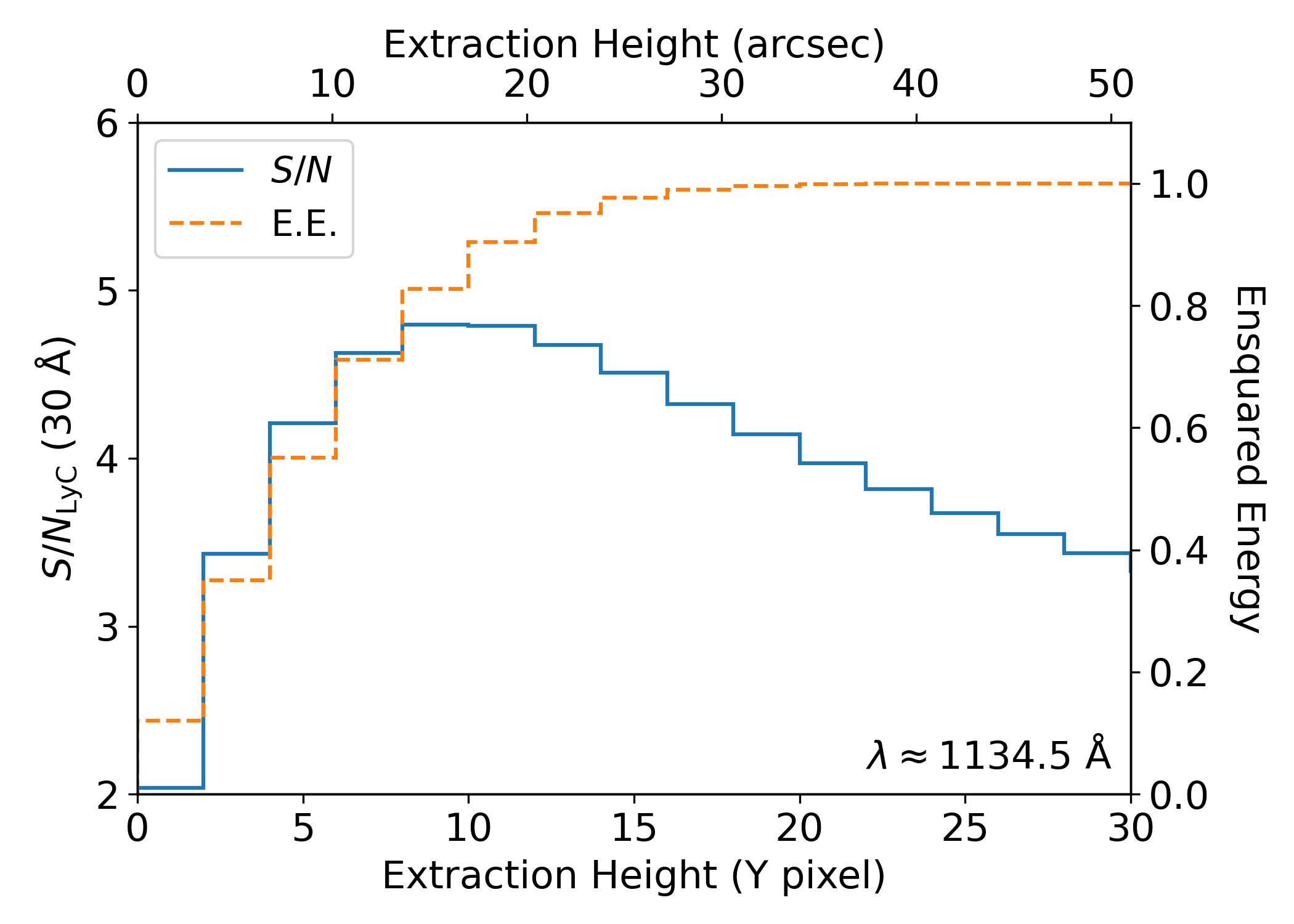}
\end{tabular}
\end{center}


\caption{Signal-to-noise ratio as a function of extraction height for the target, using the laboratory-measured emission lines N I $\lambda\lambda$1134.2, 1134.4, 1135.0. An extraction height of $\sim$~9~pixels, corresponding to $\sim$~15$^{\prime\prime}$, gives a maximum $S/N$ for point-source LyC measurements. This value is slightly wider than, but comparable to, the measured cross-dispersion FWHM of $\sim$~13$^{\prime\prime}$ shown in Fig.~\ref{fig:FitLineExample}.
\label{fig:BestSN}}
\end{figure}

We also analyze the optimal $S/N$ obtained by varying the spectral extraction height of point sources on the SPRITE MCP in Fig.~\ref{fig:BestSN}.
In this anaflysis, we use the dispersion measured from the N I $\lambda\lambda$1134.2, 1134.4, 1135.0 laboratory emission lines, representative of the observed LyC wavelengths for targets at $z > 0.2$.
We find that we achieve a maximum $S/N$ for a simulated SPRITE LyC measurement of J091113$+$183108 with an extraction height of 9 pixels, corresponding to 15 arcsec ($\approx$ 100 $\mu$m on the detector), which is slightly wider than, but comparable to, the measured cross-dispersion FWHM of $\sim$~13 arcsec discussed above. This aperture corresponds to an ensquared energy (E.E.) of 82.5\% across 30~\AA\ of Lyman continuum.
The E.E. is determined using cross-dispersion Gaussian fits to the various spectral features measured in the SPRITE laboratory analysis (Fig.~\ref{fig:FitLineExample}).
For consistency, we use J091113$+$183108 as our reference target throughout this paper because it has a detectable but moderate LyC flux, ranking fourth among the eight commissioning targets in Table~\ref{tab:TargetSummary}.

On orbit, this extraction height may be increased due to spacecraft pointing jitter.
The spacecraft pointing jitter expectation from Blue Canyon Technologies is 7$^{\prime\prime}$ RMS.
The laboratory-measured cross dispersion FWHM is $\sim$ 8 pixels (Fig.~\ref{fig:FitLineExample}), or 12.67$^{\prime\prime}$ using the detector pixel scale of 11 $\mu$m~pixel$^{-1}$ and the plate scale of 144 arcsec~mm$^{-1}$.
After subtracting the 4$^{\prime\prime}$ line-spread function of the long-tank calibration system in quadrature, the intrinsic SPRITE cross-dispersion FWHM is $\sim$12$^{\prime\prime}$.
Adding the 7$^{\prime\prime}$ jitter in quadrature gives an estimated on-orbit cross-dispersion FWHM of 13.91$^{\prime\prime}$, a $\sim$ 16\% increase.
In terms of the extraction aperture in Fig.~\ref{fig:BestSN}, this broadening increases the nominal extraction height from 9 pixels ($\approx$ 14$^{\prime\prime}$) to 10 pixels ($\approx$ 16$^{\prime\prime}$).
Pointing motion projected along the dispersion direction could in principle broaden the spectral line-spread function, but this contribution is separate from the cross-dispersion estimate used for the extraction-height calculation.
In practice, the pointing error may be smaller because the SPRITE reaction wheels are intentionally undersized to increase pointing stability.
However, we cannot estimate the pointing improvement effectively as this method does not have heritage on a Blue Canyon Technologies ADCS XACT-15 platform at the SPRITE mass.

\subsubsection{Predicted Sensitivity Relative to \textit{HST}-COS} \label{subsubsec:sensitivity_HST}

Considering the $A_{\rm{eff}}$ of the instruments and a nominal dark rate of $D_{0} =$ 4 counts cm$^{-2}$ s$^{-1}$, we compare the 3$\sigma$ sensitivity of SPRITE to that of \textit{HST}-COS G140L per resolution element, adopting $\Delta\lambda \sim$ 2~\AA\ for SPRITE and $\Delta\lambda \simeq$ 6 $\times$ 0.0803 $\approx$ 0.48~\AA\ for \textit{HST}-COS G140L [\citenum{2024cosi_book_17H}].
These calculations follow techniques similar to those employed to obtain the $S/N$ values in Section~\ref{sec:commission_plan}.
This assumes an exposure time of $t_{\rm{exp}} =$ 10$^5$ s for SPRITE, and $t_{\rm{exp}} \sim$ 4 ks for \textit{HST}-COS.
Such an exposure time corresponds to approximately 1.5--2 \textit{HST}-COS orbits, consistent with the typical per-target allocation in the LzLCS program (134 orbits for 66 targets).

\begin{figure}[htbp]
\begin{center}
\begin{tabular}{c}
\includegraphics[width=0.65\textwidth]{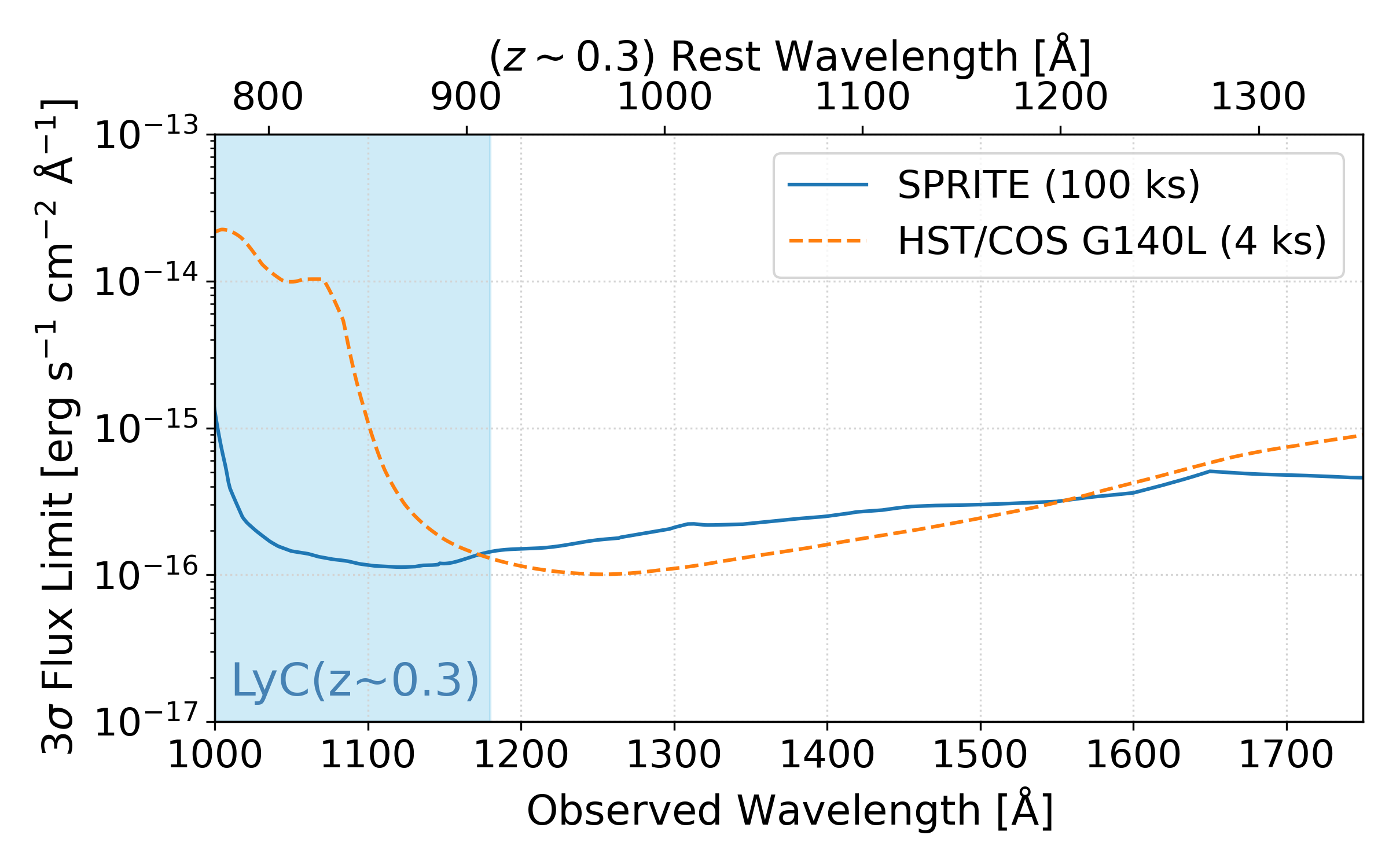}
\end{tabular}
\end{center}


\caption{3$\sigma$ continuum sensitivity of SPRITE (blue, 10$^{5}$ s, $\Delta\lambda \sim$ 2~\AA) compared to a two-orbit \textit{HST}-COS G140L exposure (orange dashed, $\Delta\lambda \sim$ 0.48~\AA) over the SPRITE's bandpass. The shaded region marks the LyC at $z \sim$ 0.3. While COS is more sensitive for most of the bandpass, SPRITE's non-competitive scheduling allows longer integrations and an order-of-magnitude higher sensitivity at around 1000--1150~\AA.
\label{fig:SPRITE_HST_Sensitivity}}
\end{figure}

In Fig.~\ref{fig:SPRITE_HST_Sensitivity}, we see that SPRITE demonstrates higher continuum sensitivity than \textit{HST}-COS at around 1000--1150~\AA, largely due to the reduced background from its smaller spectral extraction area and advanced mirror coatings.
In addition, SPRITE is an imaging spectrograph with one spatial and one spectral dimension.
Therefore, compared to other FUV spectrographs such as \textit{FUSE}, \textit{HUT}, and \textit{HST}-COS, SPRITE captures a much lower dark rate per \AA, though the grating scatter negatively impacts this metric.
Nevertheless, SPRITE's enhanced sensitivity and non-competitive scheduling enable it to explore a larger portion of the redshifted LyC bandpass below $\lambda \sim$ 1100~\AA\ ($\lambda_{\rm{rest}} \sim$ 846~\AA\ for a $z =$ 0.3 galaxy), where \textit{HST}-COS sensitivity declines rapidly and practical exposure times are severely limited.
Extending LyC studies to shorter rest-frame wavelengths than those primarily probed by LzLCS ($\lambda_{\rm{rest}} \sim$ 880--900~\AA) can provide new constraints on physical processes that shape the emergent LyC spectrum, including possible nebular free-bound emission near the Lyman edge [\citenum{2024MNRAS_530_2133S}], dust attenuation below the edge, and the intrinsic stellar ionizing spectrum predicted by population synthesis models.
Such measurements enable tests of whether LyC escape is dominated by density-bounded nebular regions, porous ionization-bounded regions, or intermediate scenarios.
This extends LyC studies beyond integrated escape fractions to the spectral properties of the escaping ionizing radiation.

Based on this analysis, SPRITE is capable of extending LyC measurements to shorter wavelengths, and repeating many of the LzLCS Lyman continuum observations, despite the increased scatter background.
As a relatively low-cost, dedicated mission with a strong focus on LyC escape, SPRITE will be able to prioritize these observations and dedicate sufficient time to reach $\sim$~10$^5$ s on each of the $\sim$~50 LyC candidate targets.

\section{SPRITE LyC Science Demonstration Plan} \label{sec:commission_plan}

SPRITE will map FUV emission from SNRs in the Magellanic Clouds and star-forming regions in nearby galaxies, as discussed in a companion paper --- Carlson et al. 2025 [\citenum{carlson2025push}].
A second pillar of the SPRITE science program --- and the focus of this paper --- is to measure or place limits on the escape fraction of LyC photons from nearby star-forming galaxies (0.16 $<z<$ 0.4).
This section outlines the commissioning plan demonstrating the SPIRES LyC survey capabilities during the first several weeks of the SPRITE mission.
These tasks and observations will demonstrate the instrument is ready to execute the LyC science program as planned.

\begin{figure}[htbp]
\begin{center}
\begin{tabular}{c}
\includegraphics[width=0.95\textwidth]{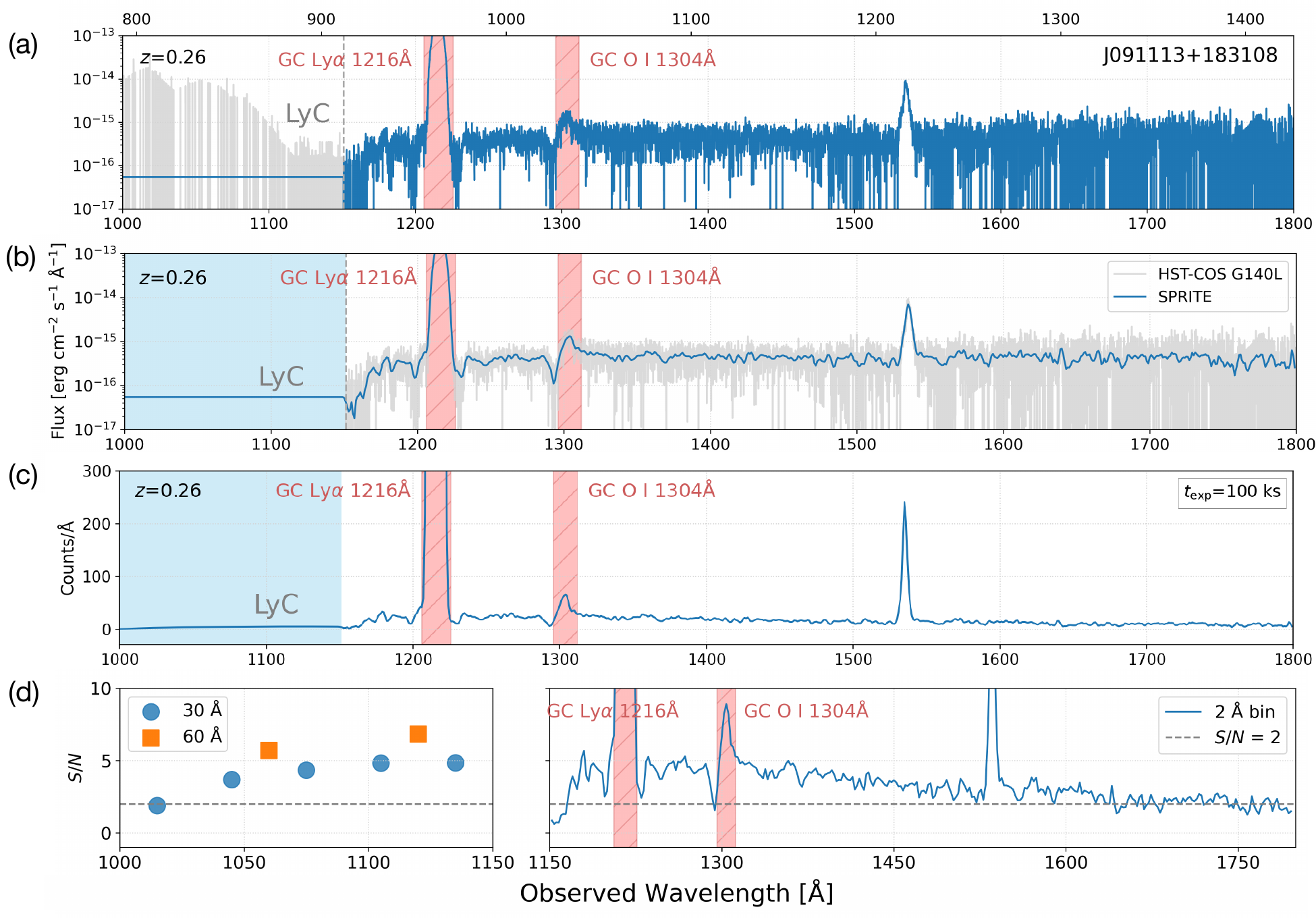}
\end{tabular}
\end{center}


\caption{Steps used to estimate $S/N$ for J091113$+$183108 with SPRITE.
(a) \textit{HST}-COS spectrum showing the LyC-averaged flux.
(b) Spectrum degraded to SPRITE's spectral FWHM ($\Delta\lambda \sim$ 2~\AA) at the observed wavelengths corresponding to $F_{1100}$.
(c) Predicted SPRITE photon counts per \AA\ using SPRITE's $A_{\rm{eff}}$ and $t_{\rm{exp}} =$ 10$^5$ s.
(d) Resulting $S/N$ for LyC (30~\AA\ and 60~\AA\ bins), $F_{1100}$ (per resolution element), and Ly$\alpha$ ($S/N_{\rm{Ly\alpha}} =$ 18.64).
The corresponding predicted $S/N$ values are listed in Table~\ref{tab:TargetSN}.}
\label{fig:procedure_flux2SN}
\end{figure}

With the LyC-averaged \textit{HST}-COS spectra of the eight commissioning targets from LzLCS shown in Fig.~\ref{fig:8targetHST}, we measure the average $F_{\rm{LyC}}$, $F_{1100}$, and $F_{\rm{Ly\alpha}}$ listed in Table \ref{tab:TargetSummary}.
To obtain the estimated signal counts, we perform the following operations, demonstrated with one of our targets --- J091113$+$183108 --- in Fig.~\ref{fig:procedure_flux2SN}.

\begin{table}[htbp]
\caption{Measured flux densities for the eight potential commissioning targets.}
\label{tab:TargetSummary}
\centering
{\footnotesize\CMtable
\vspace{5pt}
\renewcommand{\arraystretch}{1.5}
\begin{tabular}{c c c c c c c}
\hline
\noalign{\vskip 1.5pt}
\hline
Target Name &
R.A. (deg) &
Dec (deg) &
$z$ &
$F_{\rm{LyC}}$ $^{a}$ &
$F_{1100}$ $^{a}$ &
$F_{\rm{Ly\alpha}}$ $^{a}$
\\
\hline
\rule[-1ex]{0pt}{3.5ex}
J115855$+$312559 & 179.73 & 31.43 & 0.243 & 21.40 & 89.01 & 2014.32 \\
J143256$+$274249 & 218.24 & 27.71 & 0.266 & 19.78 & 88.41 & 1237.19 \\
J105331$+$523753 & 163.38 & 52.63 & 0.253 & 2.81 & 131.87 & 642.81 \\
J081409$+$211459 & 123.54 & 21.25 & 0.227 & 2.05 & 78.47 & 243.78 \\
J1442$-$0209 & 220.63 & $-$2.16 & 0.294 & 3.53 & 30.47 & 1070.82 \\
\textbf{J091113$+$183108} & 137.81 & 18.52 & 0.262 & 5.42 & 48.17 & 922.40 \\
J091703$+$315221 & 139.26 & 31.87 & 0.300 & 6.25 & 44.97 & 563.65 \\
J120934$+$305326 & 182.39 & 30.89 & 0.219 & 3.83 & 81.55 & 919.62 \\
\hline
\end{tabular}

\vspace{5pt}
\parbox{\linewidth}{\footnotesize
\textsc{Note 1} --- $F_{\rm LyC}$ is averaged from measurable flux at $\lambda \le \lambda_{\rm LyC,obs}$; $F_{1100}$ at $\lambda_{\rm F1100,obs}$; and $F_{\rm Ly\alpha}$ at $\lambda_{\rm Ly\alpha,obs}$ from \textit{HST}-COS G140L spectra in LzLCS shown in Fig.~\ref{fig:8targetHST} [\citenum{2022ApJS_260_1F}].}

\vspace{5pt}
\parbox{\linewidth}{\footnotesize
\textsc{Note 2} --- $^a$ All fluxes are in units of 10$^{-17}$ erg s$^{-1}$ cm$^{-2}$~\AA$^{-1}$.}

\vspace{5pt}
\parbox{\linewidth}{\footnotesize
\textsc{Note 3} --- J091113$+$183108 is highlighted as the representative commissioning target used in the performance analysis below.}

}
\end{table}

We degrade the \textit{HST} spectrum (G140L: $\Delta\lambda \sim$ 0.48~\AA\ [\citenum{2024cosi_book_17H}]; Fig.~\ref{fig:procedure_flux2SN}a) to the resolution of SPRITE, adopting $\Delta\lambda \sim$ 2~\AA\ near the observed wavelengths corresponding to rest-frame $F_{\rm{1100}}$ (Fig.~\ref{fig:procedure_flux2SN}b).
We derive the signal counts per \AA\ of the targets observable with SPRITE by:
\begin{equation}
S = F(\lambda) \cdot A_{\rm{eff}}(\lambda) \cdot t_{\rm{exp}} / E_{\gamma}(\lambda),
\end{equation}
where $F(\lambda)$ is the flux of the galaxy, $E_{\gamma}(\lambda) = hc/\lambda$ is the photon energy, $A_{\rm{eff}}(\lambda)$ is the light-collecting effective area of SPRITE (blue solid line in Fig.~\ref{fig:Aeff_SPRITE_Bowen}), and $t_{\rm{exp}} =$ 10$^5$ s is the estimated exposure time.
We then obtain the signal counts per \AA\ with SPRITE, as depicted in Fig.~\ref{fig:procedure_flux2SN}c.
Since the signal counts in the LyC of the spectrum is too low to be detected in a single \AA, we first bin the LyC into a size of 30~\AA.
This gives us at least three integrated data points to acquire reasonable $S/N$ for galaxies at $z >$ 0.2.
We also further bin the signal counts into 60~\AA\ for the purpose of achieving higher $S/N$.

\begin{table}[htbp]
\caption{Predicted $S/N$ for the eight potential commissioning targets.}
\label{tab:TargetSN}
\centering
{\footnotesize\CMtable
\vspace{5pt}
\renewcommand{\arraystretch}{1.5}
\begin{tabular}{c c c c c}
\hline
\noalign{\vskip 1.5pt}
\hline
Target Name &
$(S/N)_{\rm LyC}$ (30~\AA) &
$(S/N)_{\rm LyC}$ (60~\AA) &
$(S/N)_{F1100}$\textsuperscript{a} &
$(S/N)_{\rm Ly\alpha}$\textsuperscript{a} \\
\hline
\rule[-1ex]{0pt}{3.5ex}%
J115855$+$312559 & 14.64 & 21.32 & 6.37 & 23.46 \\
J143256$+$274249 & 14.47 & 20.40 & 6.01 & 18.94 \\
J105331$+$523753 & 2.55  & 3.68  & 7.76 & 15.26 \\
J081409$+$211459 & 1.73  & 2.59  & 6.11 & 7.23  \\
J1442$-$0209     & 3.27  & 4.40  & 2.66 & 17.83 \\
\textbf{J091113$+$183108} & 4.82  & 6.84  & 3.94 & 18.64 \\
J091703$+$315221 & 5.46  & 7.28  & 3.55 & 11.85 \\
J120934$+$305326 & 3.04  & 4.56  & 6.43 & 19.85 \\
\hline
\end{tabular}

\vspace{2mm}
\parbox{\linewidth}{\footnotesize
\textsc{Note 1} --- Assuming a nominal dark rate of $D_{0} = 4$ counts cm$^{-2}$ s$^{-1}$, $A_{\rm{eff}}$ of SPRITE and $t_{\rm{exp}} = 10^5$ s. $^a$ $F_{1100}$ and Ly$\alpha$ counts are shown per resolution element ($\sim 2$~\AA).}

\vspace{2mm}
\parbox{\linewidth}{\footnotesize
\textsc{Note 2} --- SPRITE is expected to readily detect LyC emission in 60~\AA\ bin} with $S/N > 3$ for all but J081409$+$211459 due to its extremely low $F_{\rm{LyC}}$.

\vspace{5pt}
\parbox{\linewidth}{\footnotesize
\textsc{Note 3} --- J091113$+$183108 is highlighted as the representative commissioning target used in the performance analysis below.}

}
\end{table}

With a nominal dark rate of $D_{0} =$ 4 counts cm$^{-2}$ s$^{-1}$, we calculate the $S/N$ by:
\begin{equation}
S/N = \frac{S}{\sqrt{S+D}},
\end{equation}
where $S$ is the total detected number of photons, and $D$ is the dark counts including grating scatter (see \S~\ref{subsec:SPRITE_gratingscatter}).
For J091113$+$183108, we acquire $S/N$ of $\sim$~5 with 30~\AA-binned LyC counts ($\sim$~7 for 60~\AA-binned), as well as the non-ionizing FUV spectrum at $F_{1100}$ shown as per resolution element (Fig.~\ref{fig:procedure_flux2SN}d).
Our best initial estimates of $S/N$ for all of our commissioning targets are shown in Table \ref{tab:TargetSN}.
We stress that the LyC are integrated with a bin size of 30 and 60~\AA\ for reasonably detectable $S/N$.
For targets with $S/N <$ 3 in $F_{1100}$, we will also bin the spectra accordingly to obtain meaningful measurements.

\section{Performance Margins} \label{sec:degrade}

As a SmallSat with significant technology demonstration, it is important that the SPRITE mission carries margins in its science performance.
In this section, investigate the impacts of degraded effective area and various background levels on these margins.

We also consider the impact of excess spacecraft jitter.
As discussed in Section~\ref{subsec:SPRITE_sensitivity}, the nominal 7$^{\prime\prime}$ RMS pointing-jitter requirement broadens the estimated on-orbit cross dispersion FWHM to 13.91$^{\prime\prime}$.
If the jitter increases to 10$^{\prime\prime}$ RMS, the estimated cross-dispersion FWHM would become 15.6$^{\prime\prime}$, corresponding to an extraction height of approximately 10 pixels.
The primary impact would be increased background within the extraction aperture and a modest decrease in $S/N$.
Jitter projected along the dispersion direction could in principle broaden the spectral line-spread function, but this effect is expected to be secondary for the broad 30--60~\AA\ LyC bins used here.

\subsection{Degraded effective area but same background} \label{subsec:degradeAeff_sameBg}

Environmental factors, such as variations in temperature, radiation, or contamination, will cause the mirror's reflectivity to degrade over time.
The reflectance of the flight coatings on the mirrors may also differ from values measured on witness samples, and may further be affected by humidity exposure in the final weeks before launch, as SPRITE cannot be stored in a dry environment after handover to the launch service provider. 

As the dominant noise source is scatter within the instrument, a decrease in throughput should produce a corresponding decrease in noise.
However, for this analysis, we leave the background rate constant.
We take the nominal effective area, $A_{\rm{eff,0}}$, shown in Fig.~\ref{fig:Aeff_SPRITE_Bowen}, as the baseline throughput for these comparisons.

\begin{figure*}[htbp]
    \centering
    \includegraphics[width=\textwidth]{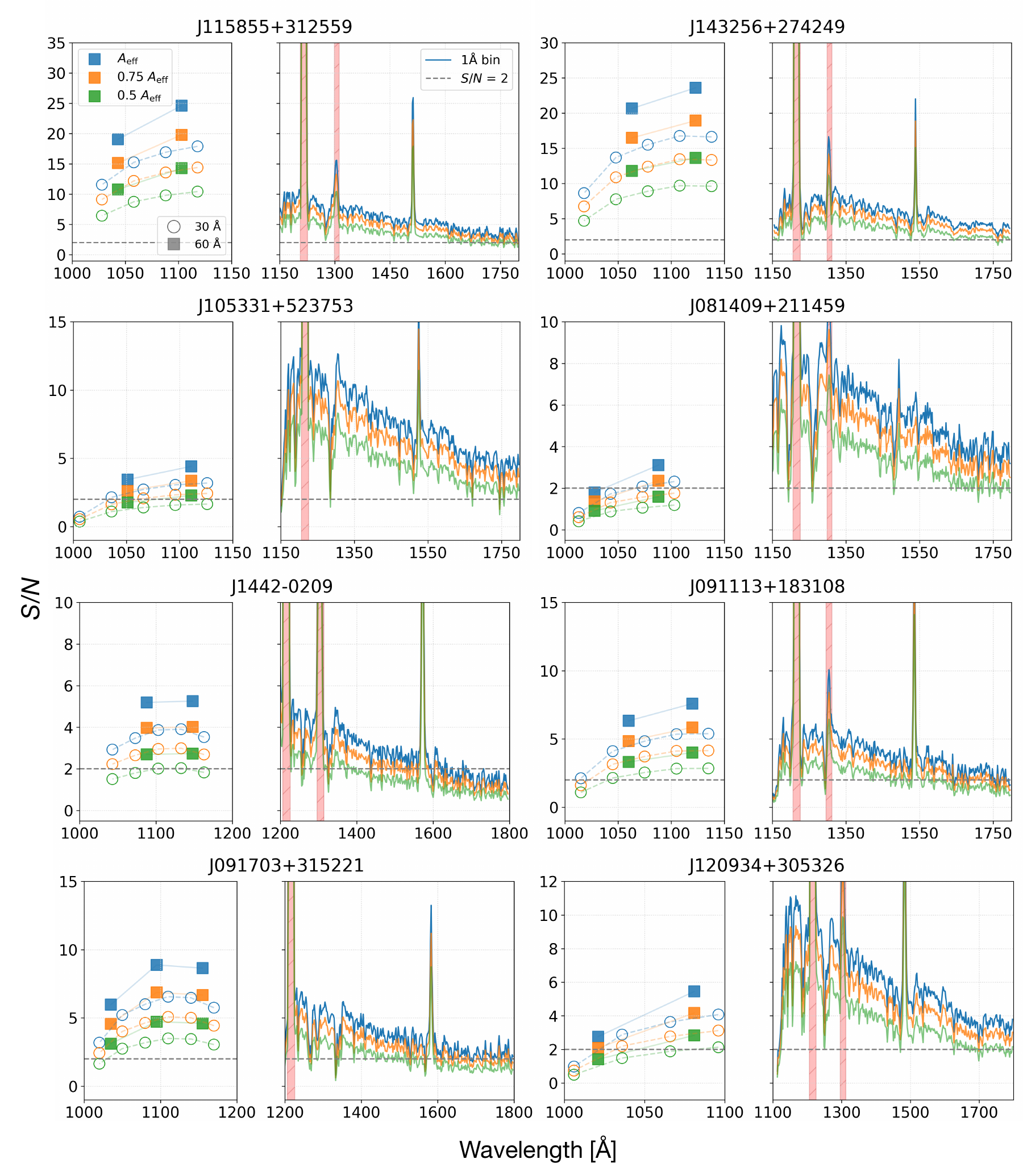}

    \caption{$S/N$ predictions for the eight commissioning targets under nominal SPRITE performance (blue) and effective area degradations to 0.75 $A_{\rm{eff},0}$ (orange), and 0.5 $A_{\rm{eff},0}$ (green). 
    Despite the degradation, all targets, except J081409$+$211459, retain detectable LyC at $S/N >$ 2 using 60~\AA\ binning.
    We do not show the Ly$\alpha$ peaks here to focus on the low $S/N$ of LyC.
    \label{fig:VaryAeff}}
\end{figure*}

In Fig.~\ref{fig:VaryAeff}, we illustrate that even with a degraded $A_{\rm{eff}}$ of 0.75 $A_{\rm{eff,0}}$, we can still achieve $S/N >$~2 for the integrated LyC of all our commissioning targets using a 60~\AA\ binning.
We will be able to obtain $S/N >$ 2 for all targets but J081409$+$211459, even if $A_{\rm{eff,0}}$ is reduced to half.

Recent XeLiF coating tests show only modest FUV degradation of $\sim$1--2\% after storage at 40\% relative humidity for 3.5 years [\citenum{2025JATIS_11d2209Q}].
Therefore, the 25\% reduction in $A_{\rm eff}$ considered here should be interpreted as a conservative stress test of the total system throughput.
It includes coating degradation, contamination, alignment changes, or other storage and launch effects, rather than the expected degradation of the XeLiF-coated grating alone.

\subsection{Higher background but similar effective area} \label{subsec:higherBg_sameAeff}

\begin{figure*}[htbp]
    \centering
    \includegraphics[width=0.99\textwidth]{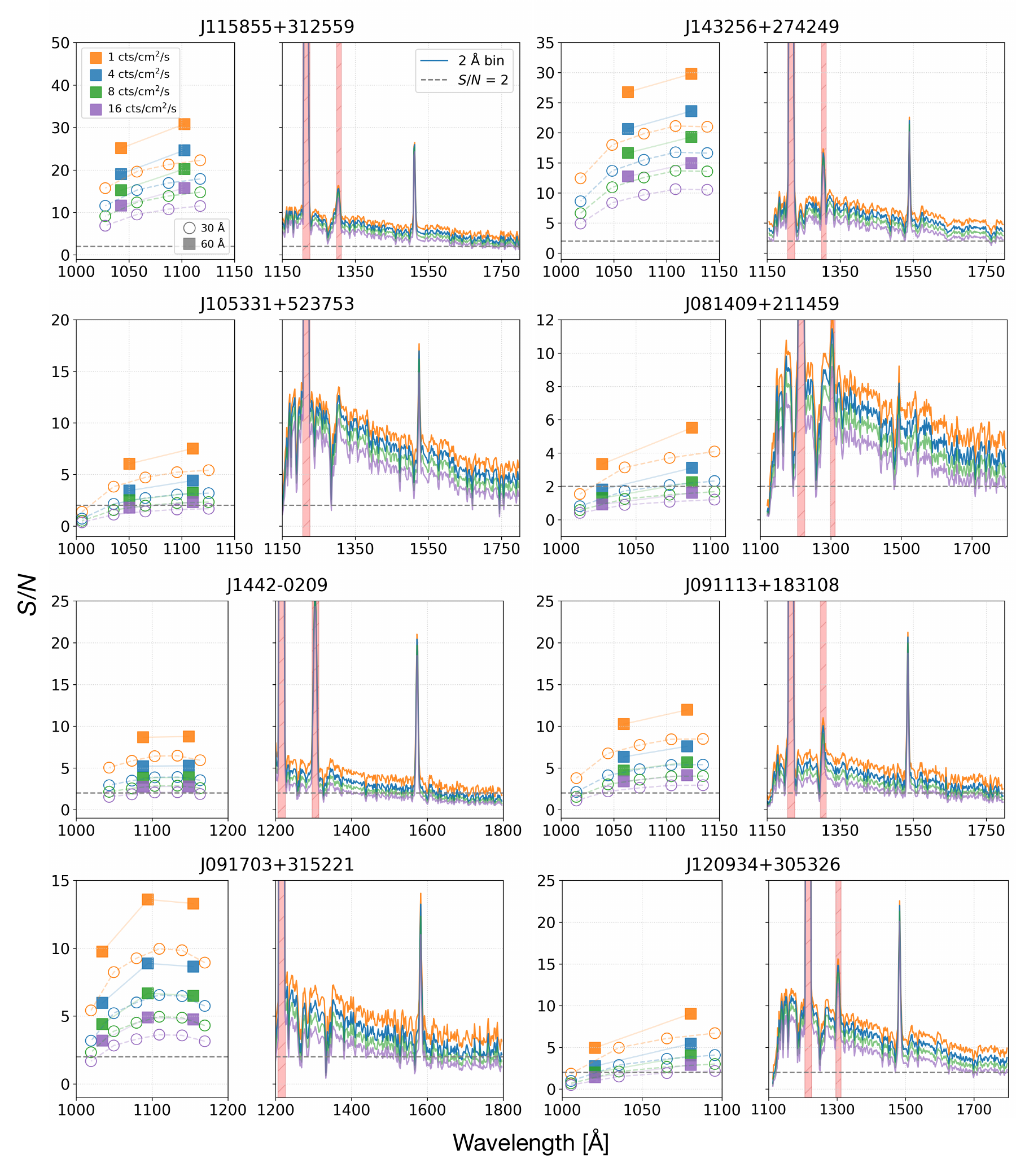}
    \caption{$S/N$ predictions for varying dark rates: $D$ = 4 (nominal; blue), 1 (orange), 8 (green), and 16 (purple) counts cm$^{-2}$ s$^{-1}$. All targets except J081409$+$211459 will achieve $S/N >$ 2 even under a fourfold background increase.
    \label{fig:VaryDarkCts}}
\end{figure*}

We begin with our nominal dark rate of $D_{0} =$ 4 counts cm$^{-2}$ s$^{-1}$ and explore scenarios with noise levels at 0.5, 2, and 4 times the nominal background rate.
These cases assume that the mirrors are in their perfect conditions, with $A_{\rm{eff}} = A_{\rm{eff},0}$.
In Fig.~\ref{fig:VaryDarkCts}, all targets except for J081409$+$211459 acquire $S/N >$ 2 in the LyC if noise increases to 4 times the nominal rate.

\subsection{Degraded effective area and higher background} \label{subsec:degradeAeff_higherBg}

\begin{figure*}[htbp]
    \centering
    \includegraphics[width=0.95\textwidth]{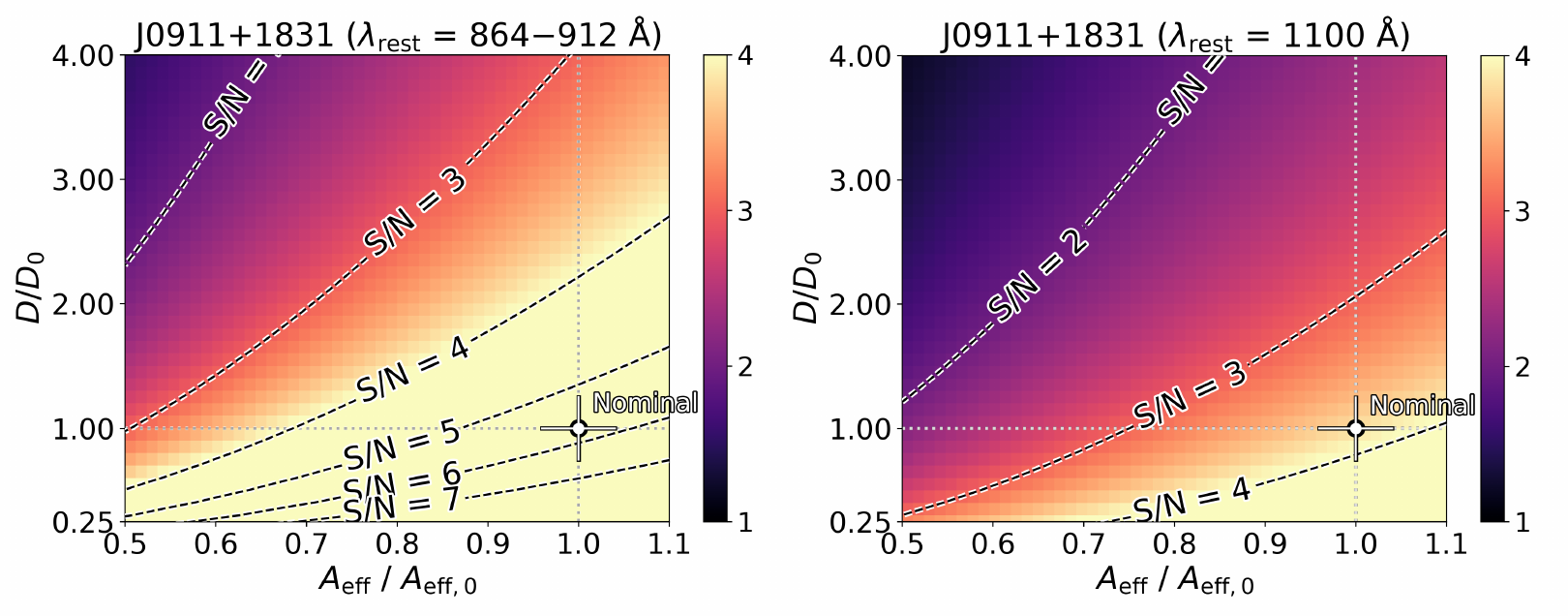}
    
    \caption{Combined effects of $A_{\rm{eff}}$ degradation and background level on the predicted $S/N$ for J091113$+$183108. Left: LyC (60~\AA-binned, $\lambda_{\rm rest}=864$--912~\AA). Right: $F_{1100}$. Contours show constant $S/N$ across a grid of $A_{\rm{eff}}/A_{\rm{eff},0}$ and $D/D_0$. The nominal operating condition is marked as a cross. SPRITE achieves $S/N \gtrsim 2$ for LyC and $F_{1100}$ over a wide range of degradation combinations.
    \label{fig:VaryAeffAndBg}}
\end{figure*}

In space, SPRITE will encounter situations where the mirror degrades and the telescope detects a background level that differs from our nominal value.
In such cases, we examine how the $S/N$ for LyC and $F_{1100}$ varies using J091113$+$183108 as a representative target (Fig.~\ref{fig:VaryAeffAndBg}).
We choose this target because it has a detectable LyC flux, but not one of the brightest.
With dark rates varying from 0.5 to 4 times the nominal value, and $A_{\rm{eff}}$ from 0.5 to 1.1$\times A_{\rm{eff,0}}$, we show that we are able to achieve $S/N \sim$ 1--7 with integrated LyC (30 and 60~\AA).
For $F_{1100}$, we obtain $S/N \sim$ 1--4 in SPRITE's spectral resolution ($\sim$~2~\AA).
When needed, we are able to bin further to obtain higher $S/N$ for robust measurements.

\begin{table}[htbp]
\caption{Signal-To-Noise Ratios of J091113$+$183108 with Degraded Factors.} 
\label{tab:J091113_183108_Vary}
\centering
{\footnotesize\CMtable
\vspace{5pt}
\renewcommand{\arraystretch}{1.5}
\begin{tabular}{c c c c c c}
\hline
\noalign{\vskip 1.5pt}
\hline
$A_{\rm{eff}}/A_{\rm{eff,0}}$ &
$D/D_0$ &
$(S/N)_{\rm{LyC}}$ (30~\AA) &
$(S/N)_{\rm{LyC}}$ (60~\AA) &
$(S/N)_{\rm{F1100}}$ &
$(S/N)_{\rm{Ly\alpha}}$ \\
\hline\hline
\rule[-1ex]{0pt}{3.5ex}
1.00 & 1 & 4.82 & 6.84 & 3.94 & 18.64 \\
\rule[-1ex]{0pt}{3.5ex}
1.00 & 2 & 3.58 & 5.08 & 3.19 & 17.88 \\
\rule[-1ex]{0pt}{3.5ex}
1.00 & 4 & 2.60 & 3.69 & 2.46 & 16.60 \\
\rule[-1ex]{0pt}{3.5ex}
0.75 & 1 & 3.70 & 5.25 & 3.15 & 15.92 \\
\rule[-1ex]{0pt}{3.5ex}
0.50 & 1 & 2.53 & 3.59 & 2.26 & 12.64 \\
\rule[-1ex]{0pt}{3.5ex}
0.75 & 2 & 2.72 & 3.86 & 2.49 & 15.09 \\
\rule[-1ex]{0pt}{3.5ex}
0.50 & 2 & 1.84 & 2.61 & 1.74 & 11.74 \\
\rule[-1ex]{0pt}{3.5ex}
0.75 & 4 & 1.96 & 2.79 & 1.89 & 13.76 \\
\rule[-1ex]{0pt}{3.5ex}
0.50 & 4 & 1.32 & 1.87 & 1.29 & 10.39 \\
\hline

\end{tabular}

\vspace{5pt}
\parbox{\linewidth}{\footnotesize
\textsc{Note} --- Nominal example of possible degrade function of SPRITE using the \textit{HST}-COS spectrum of J091113$+$183108, with nominal background of $D_0$ = 4 counts cm$^{-2}$ s$^{-1}$, and $A_{\rm{eff},0}$ shown in Fig.~\ref{fig:Aeff_SPRITE_Bowen}.}
}
\end{table}

We summarize the expected $S/N$ with different combinations of degraded $A_{\rm{eff}}$ and increased background in Table \ref{tab:J091113_183108_Vary}.
Across the degradation combination, we are able to obtain $S/N \gtrsim$~2 in LyC, $F_{1100}$ when the fluxes are integrated, and better for Ly$\alpha$ emission due to its bright nature.

\section{Summary and Advances in Longer Bandpass Observations} \label{sec:summary}

This work demonstrates SPRITE's capability to detect low-redshift LyC galaxies (0.16 $<z<$ 0.4).
With advanced Lyman-UV optical coatings, including XeLiF on the grating, SPRITE achieves an $A_{\rm{eff}}$ comparable to large FUV space telescopes such as \textit{HUT} and \textit{FUSE} despite its compact size.
Because of the small spectral extraction area of an imaging spectrograph, SPRITE detects a lower dark rate per \AA\ compared with the non-imaging spectrographs such as \textit{HUT}, \textit{FUSE}, and \textit{HST}-COS.
These advancements enable SPRITE to reach a sensitivity flux limit of $\sim$~10$^{-16}$ to 10$^{-15}$ erg s$^{-1}$ cm$^{-2}$~\AA$^{-1}$ at 1000 $<\lambda<$ 1150~\AA, which covers the rest wavelength span of $\gtrsim$ 100~\AA\ of LyC observed in the redshifted galaxies at $z \sim$ 0.3.
This also makes SPRITE more sensitive than \textit{HST}-COS for detecting LyC of such galaxies.

We select eight commissioning targets from LzLCS to demonstrate that SPRITE is able to detect LyC emission from known LyC leakers in the $z \sim$ 0.3 range.
Based on this analysis, SPRITE should be capable of measuring an expanded sample of unknown galaxies at flux levels similar to those in the LzLCS sample.
The definition of this new SPRITE sample represents the continuation of this work, carried out in collaboration with the SPRITE science team, which includes several LzLCS team members, including the LzLCS PI.

The commissioning targets are selected from the LzLCS sample based on the following criteria:
(1) highest LyC flux,
(2) highest flux at $\lambda_{\rm{rest}}$ = 1100~\AA,
(3) highest Ly$\alpha$ flux, and
(4) observable LyC binned to 60~\AA\ with SPRITE ($S/N >$ 4).
We show that SPRITE is able to detect the commissioning targets by integration of LyC spectra, despite grating scatter and small size, due to its efficient optical design and HWO-enabling technology.
This demonstrates that SPRITE is capable of measuring LyC photons from low-redshift LCEs.

We also inspect SPRITE's performance margin under degraded mirror coatings and varying background levels.
For example, using one of our median commissioning targets, J091113$+$183108, we are able to achieve $S/N >$ 2 in $F_{\rm{LyC}}$ and $F_{1100}$ when the fluxes are integrated, across different combinations of degraded $A_{\rm{eff}}$ and background.

Future work will focus on identifying a broader sample of LyC-emitting galaxies through archival data mining using SDSS and \textit{GALEX}, as well as cross-comparison with theoretical models such as Starburst99.
Once SPRITE obtains spectra of these targets, we will measure their LyC and Ly$\alpha$ escape fractions, [O~III]/[O~II] ratios, H$\beta$ equivalent width, and related diagnostics.
These measurements will allow for a more comprehensive exploration of LyC escape mechanisms at low redshifts (0.16 $< z <$ 0.4), which can serve as proxies for the high-redshift galaxies observed by \textit{JWST}.

SPRITE's ability to probe the LyC spectral shape at wavelengths inaccessible to \textit{HST}-COS provides new constraints on ionizing photon escape, including the relative contributions of stellar versus nebular emission and dust attenuation beyond the Lyman limit.
Ultimately, SPRITE will play a key role in advancing our understanding of the physical processes that govern ionizing photon escape and improving constraints on how galaxies contributed to cosmic reionization.
SPRITE's on-orbit Lyman-UV performance will also inform the design of future missions such as HWO.
The mission is scheduled to launch in October 2026.

\subsection* {Disclosures}

The authors declare that there are no financial interests, commercial affiliations, or other potential conflicts of interest that could have influenced the objectivity of this research or the writing of this paper.

\subsection* {Code and Data Availability}

The data supporting the findings of this article are proprietary and are not publicly available, but can be replicated via the published code.
The archived version of the code described in this paper can be freely accessed through GitHub via the SPRITE data reduction pipeline.

\subsection* {Acknowledgments}

This work was funded by a grant from the National Aeronautics and Space Administration (NASA) (Grant Nos. 80NSSC19K0995 and 80NSSC24K0231) to the University of Colorado, Boulder.
The authors acknowledge the use of \textit{Hubble Space Telescope} --- Cosmic Origins Spectrograph (\textit{HST}-COS) data retrieved from the Mikulski Archive for Space Telescopes (MAST) at the Space Telescope Science Institute (STScI).
We thank Alex Haughton and Celeste Elizalde Flores for helpful paper discussions.
We also extend many thanks to all the staff and students at LASP who have contributed their time and
effort to the project.
The authors acknowledge the use of grammarly and ChatGPT (OpenAI) to assist language and grammar refinement.
All AI-generated suggestions were reviewed, edited, and approved by the authors to ensure accuracy and integrity.


\bibliography{report}   

@ARTICLE{2022ApJS_260_1F,
       author = {{Flury}, Sophia R. and {Jaskot}, Anne E. and {Ferguson}, Harry C. and {Worseck}, G{\'a}bor and {Makan}, Kirill and {Chisholm}, John and {Saldana-Lopez}, Alberto and {Schaerer}, Daniel and {McCandliss}, Stephan and {Wang}, Bingjie and {Ford}, N.~M. and {Heckman}, Timothy and {Ji}, Zhiyuan and {Giavalisco}, Mauro and {Amorin}, Ricardo and {Atek}, Hakim and {Blaizot}, Jeremy and {Borthakur}, Sanchayeeta and {Carr}, Cody and {Castellano}, Marco and {Cristiani}, Stefano and {De Barros}, Stephane and {Dickinson}, Mark and {Finkelstein}, Steven L. and {Fleming}, Brian and {Fontanot}, Fabio and {Garel}, Thibault and {Grazian}, Andrea and {Hayes}, Matthew and {Henry}, Alaina and {Mauerhofer}, Valentin and {Micheva}, Genoveva and {Oey}, M.~S. and {Ostlin}, Goran and {Papovich}, Casey and {Pentericci}, Laura and {Ravindranath}, Swara and {Rosdahl}, Joakim and {Rutkowski}, Michael and {Santini}, Paola and {Scarlata}, Claudia and {Teplitz}, Harry and {Thuan}, Trinh and {Trebitsch}, Maxime and {Vanzella}, Eros and {Verhamme}, Anne and {Xu}, Xinfeng},
        title = "{The Low-redshift Lyman Continuum Survey. I. New, Diverse Local Lyman Continuum Emitters}",
      journal = {\apjs},
         year = 2022,
        month = may,
       volume = {260},
       number = {1},
          eid = {1},
        pages = {1},
          doi = {10.3847/1538-4365/ac5331},
archivePrefix = {arXiv},
       eprint = {2201.11716},
 primaryClass = {astro-ph.GA},
       adsurl = {https://ui.adsabs.harvard.edu/abs/2022ApJS..260....1F}
}

@ARTICLE{2022ApJ_930_126F,
       author = {{Flury}, Sophia R. and {Jaskot}, Anne E. and {Ferguson}, Harry C. and {Worseck}, G{\'a}bor and {Makan}, Kirill and {Chisholm}, John and {Saldana-Lopez}, Alberto and {Schaerer}, Daniel and {McCandliss}, Stephan R. and {Xu}, Xinfeng and {Wang}, Bingjie and {Oey}, M.~S. and {Ford}, N.~M. and {Heckman}, Timothy and {Ji}, Zhiyuan and {Giavalisco}, Mauro and {Amor{\'\i}n}, Ricardo and {Atek}, Hakim and {Blaizot}, Jeremy and {Borthakur}, Sanchayeeta and {Carr}, Cody and {Castellano}, Marco and {De Barros}, Stephane and {Dickinson}, Mark and {Finkelstein}, Steven L. and {Fleming}, Brian and {Fontanot}, Fabio and {Garel}, Thibault and {Grazian}, Andrea and {Hayes}, Matthew and {Henry}, Alaina and {Mauerhofer}, Valentin and {Micheva}, Genoveva and {Ostlin}, Goran and {Papovich}, Casey and {Pentericci}, Laura and {Ravindranath}, Swara and {Rosdahl}, Joakim and {Rutkowski}, Michael and {Santini}, Paola and {Scarlata}, Claudia and {Teplitz}, Harry and {Thuan}, Trinh and {Trebitsch}, Maxime and {Vanzella}, Eros and {Verhamme}, Anne},
        title = "{The Low-redshift Lyman Continuum Survey. II. New Insights into LyC Diagnostics}",
      journal = {\apj},
         year = 2022,
        month = may,
       volume = {930},
       number = {2},
          eid = {126},
        pages = {126},
          doi = {10.3847/1538-4357/ac61e4},
archivePrefix = {arXiv},
       eprint = {2203.15649},
 primaryClass = {astro-ph.GA},
       adsurl = {https://ui.adsabs.harvard.edu/abs/2022ApJ...930..126F}
}

@INPROCEEDINGS{2017SPIE10401E_19H,
       author = {{Hennessy}, John and {Moore}, Christopher S. and {Balasubramanian}, Kunjithapatham and {Jewell}, April D. and {Carter}, Christian and {France}, Kevin and {Nikzad}, Shouleh},
        title = "{Atomic layer deposition and etching methods for far ultraviolet aluminum mirrors}",
    booktitle = {Society of Photo-Optical Instrumentation Engineers (SPIE) Conference Series},
         year = 2017,
       series = {Society of Photo-Optical Instrumentation Engineers (SPIE) Conference Series},
       volume = {10401},
        month = sep,
          eid = {1040119},
        pages = {1040119},
          doi = {10.1117/12.2274633},
       adsurl = {https://ui.adsabs.harvard.edu/abs/2017SPIE10401E..19H}
}

@INPROCEEDINGS{2023SPIE12678E_0AB,
       author = {{Bowen}, Maitland and {Fleming}, Brian and {Indahl}, Briana and {Vorobiev}, Dmitry and {Szewczyk}, Daniel and {France}, Kevin and {Rodr{\'\i}guez-de Marcos}, Luis V. and {Quijada}, Manuel A. and {Hennessey}, John J.},
        title = "{Preliminary optical performance of the SPRITE CubeSat instrument}",
    booktitle = {UV, X-Ray, and Gamma-Ray Space Instrumentation for Astronomy XXIII},
         year = 2023,
       editor = {{Siegmund}, Oswald H. and {Hoadley}, Keri},
       series = {Society of Photo-Optical Instrumentation Engineers (SPIE) Conference Series},
       volume = {12678},
        month = oct,
          eid = {126780A},
        pages = {126780A},
          doi = {10.1117/12.2677668},
       adsurl = {https://ui.adsabs.harvard.edu/abs/2023SPIE12678E..0AB}
}

@INPROCEEDINGS{2024SPIE13093E_34B,
       author = {{Bowen}, Maitland and {Fleming}, Brian and {Indahl}, Briana and {Vorobiev}, Dmitry and {O'Sullivan}, D{\'o}nal and {France}, Kevin and {Snyder}, Will and {Ochoa}, Alan and {Rodr{\'\i}guez-de Marcos}, Luis V. and {Quijada}, Manuel A. and {Hennessy}, John J. and {Siegmund}, Oswald H. and {Martin}, Adrian},
        title = "{Preflight characterization of the SPRITE CubeSat: a far-UV imaging spectrograph for stellar feedback in local galaxies}",
    booktitle = {Space Telescopes and Instrumentation 2024: Ultraviolet to Gamma Ray},
         year = 2024,
       editor = {{den Herder}, Jan-Willem A. and {Nikzad}, Shouleh and {Nakazawa}, Kazuhiro},
       series = {Society of Photo-Optical Instrumentation Engineers (SPIE) Conference Series},
       volume = {13093},
        month = aug,
          eid = {1309334},
        pages = {1309334},
          doi = {10.1117/12.3020453},
       adsurl = {https://ui.adsabs.harvard.edu/abs/2024SPIE13093E..34B}
}

@ARTICLE{2012ApJ_744_60G,
       author = {{Green}, James C. and {Froning}, Cynthia S. and {Osterman}, Steve and {Ebbets}, Dennis and {Heap}, Sara H. and {Leitherer}, Claus and {Linsky}, Jeffrey L. and {Savage}, Blair D. and {Sembach}, Kenneth and {Shull}, J. Michael and {Siegmund}, Oswald H.~W. and {Snow}, Theodore P. and {Spencer}, John and {Stern}, S. Alan and {Stocke}, John and {Welsh}, Barry and {B{\'e}land}, St{\'e}phane and {Burgh}, Eric B. and {Danforth}, Charles and {France}, Kevin and {Keeney}, Brian and {McPhate}, Jason and {Penton}, Steven V. and {Andrews}, John and {Brownsberger}, Kenneth and {Morse}, Jon and {Wilkinson}, Erik},
        title = "{The Cosmic Origins Spectrograph}",
      journal = {\apj},
         year = 2012,
        month = jan,
       volume = {744},
       number = {1},
          eid = {60},
        pages = {60},
          doi = {10.1088/0004-637X/744/1/6010.1086/141956},
archivePrefix = {arXiv},
       eprint = {1110.0462},
 primaryClass = {astro-ph.IM},
       adsurl = {https://ui.adsabs.harvard.edu/abs/2012ApJ...744...60G}
}

@INCOLLECTION{2024cosi_book_17H,
       author = {{Hirschauer}, A.~S.},
        title = "{COS Instrument Handbook v. 17.0}",
    booktitle = {COS Instrument Handbook v. 17.0},
         year = 2024,
       volume = {17},
        pages = {17},
       adsurl = {https://ui.adsabs.harvard.edu/abs/2024cosi.book...17H}
}

@ARTICLE{2017ApJ_845_111M,
       author = {{McCandliss}, Stephan R. and {O'Meara}, John M.},
        title = "{Flux Sensitivity Requirements for the Detection of Lyman Continuum Radiation Drop-ins from Star-forming Galaxies below Redshifts of 3}",
      journal = {\apj},
         year = 2017,
        month = aug,
       volume = {845},
       number = {2},
          eid = {111},
        pages = {111},
          doi = {10.3847/1538-4357/aa7fbb},
archivePrefix = {arXiv},
       eprint = {1707.03880},
 primaryClass = {astro-ph.GA},
       adsurl = {https://ui.adsabs.harvard.edu/abs/2017ApJ...845..111M}
}

@ARTICLE{1995ApJ_454L_19L,
       author = {{Leitherer}, Claus and {Ferguson}, Henry C. and {Heckman}, Timothy M. and {Lowenthal}, James D.},
        title = "{The Lyman Continuum in Starburst Galaxies Observed with the Hopkins Ultraviolet Telescope}",
      journal = {\apjl},
         year = 1995,
        month = nov,
       volume = {454},
        pages = {L19},
          doi = {10.1086/309760},
       adsurl = {https://ui.adsabs.harvard.edu/abs/1995ApJ...454L..19L}
}

@ARTICLE{2006AA_448_513B,
       author = {{Bergvall}, N. and {Zackrisson}, E. and {Andersson}, B. -G. and {Arnberg}, D. and {Masegosa}, J. and {{\"O}stlin}, G.},
        title = "{First detection of Lyman continuum escape from a local starburst galaxy. I. Observations of the luminous blue compact galaxy Haro 11 with the Far Ultraviolet Spectroscopic Explorer (FUSE)}",
      journal = {\aap},
         year = 2006,
        month = mar,
       volume = {448},
       number = {2},
        pages = {513-524},
          doi = {10.1051/0004-6361:20053788},
       adsurl = {https://ui.adsabs.harvard.edu/abs/2006A&A...448..513B}
}

@ARTICLE{2011AA_532A_107L,
       author = {{Leitet}, E. and {Bergvall}, N. and {Piskunov}, N. and {Andersson}, B. -G.},
        title = "{Analyzing low signal-to-noise FUSE spectra. Confirmation of Lyman continuum escape from Haro 11}",
      journal = {\aap},
         year = 2011,
        month = aug,
       volume = {532},
          eid = {A107},
        pages = {A107},
          doi = {10.1051/0004-6361/201015654},
archivePrefix = {arXiv},
       eprint = {1106.1178},
 primaryClass = {astro-ph.CO},
       adsurl = {https://ui.adsabs.harvard.edu/abs/2011A&A...532A.107L}
}

@ARTICLE{2013AA_553A_106L,
       author = {{Leitet}, E. and {Bergvall}, N. and {Hayes}, M. and {Linn{\'e}}, S. and {Zackrisson}, E.},
        title = "{Escape of Lyman continuum radiation from local galaxies. Detection of leakage from the young starburst Tol 1247-232}",
      journal = {\aap},
         year = 2013,
        month = may,
       volume = {553},
          eid = {A106},
        pages = {A106},
          doi = {10.1051/0004-6361/201118370},
archivePrefix = {arXiv},
       eprint = {1302.6971},
 primaryClass = {astro-ph.CO},
       adsurl = {https://ui.adsabs.harvard.edu/abs/2013A&A...553A.106L}
}

@ARTICLE{2016Natur_529_178I,
       author = {{Izotov}, Y.~I. and {Orlitov{\'a}}, I. and {Schaerer}, D. and {Thuan}, T.~X. and {Verhamme}, A. and {Guseva}, N.~G. and {Worseck}, G.},
        title = "{Eight per cent leakage of Lyman continuum photons from a compact, star-forming dwarf galaxy}",
      journal = {\nat},
         year = 2016,
        month = jan,
       volume = {529},
       number = {7585},
        pages = {178-180},
          doi = {10.1038/nature16456},
archivePrefix = {arXiv},
       eprint = {1601.03068},
 primaryClass = {astro-ph.GA},
       adsurl = {https://ui.adsabs.harvard.edu/abs/2016Natur.529..178I}
}

@ARTICLE{2016MNRAS_461_3683I,
       author = {{Izotov}, Y.~I. and {Schaerer}, D. and {Thuan}, T.~X. and {Worseck}, G. and {Guseva}, N.~G. and {Orlitov{\'a}}, I. and {Verhamme}, A.},
        title = "{Detection of high Lyman continuum leakage from four low-redshift compact star-forming galaxies}",
      journal = {\mnras},
         year = 2016,
        month = oct,
       volume = {461},
       number = {4},
        pages = {3683-3701},
          doi = {10.1093/mnras/stw1205},
archivePrefix = {arXiv},
       eprint = {1605.05160},
 primaryClass = {astro-ph.GA},
       adsurl = {https://ui.adsabs.harvard.edu/abs/2016MNRAS.461.3683I}
}

@ARTICLE{2018MNRAS_474_4514I,
       author = {{Izotov}, Y.~I. and {Schaerer}, D. and {Worseck}, G. and {Guseva}, N.~G. and {Thuan}, T.~X. and {Verhamme}, A. and {Orlitov{\'a}}, I. and {Fricke}, K.~J.},
        title = "{J1154+2443: a low-redshift compact star-forming galaxy with a 46 per cent leakage of Lyman continuum photons}",
      journal = {\mnras},
         year = 2018,
        month = mar,
       volume = {474},
       number = {4},
        pages = {4514-4527},
          doi = {10.1093/mnras/stx3115},
archivePrefix = {arXiv},
       eprint = {1711.11449},
 primaryClass = {astro-ph.GA},
       adsurl = {https://ui.adsabs.harvard.edu/abs/2018MNRAS.474.4514I}
}

@ARTICLE{2018MNRAS_478_4851I,
       author = {{Izotov}, Y.~I. and {Worseck}, G. and {Schaerer}, D. and {Guseva}, N.~G. and {Thuan}, T.~X. and {Fricke}, Verhamme, A. and {Orlitov{\'a}}, I.},
        title = "{Low-redshift Lyman continuum leaking galaxies with high [O III]/[O II] ratios}",
      journal = {\mnras},
         year = 2018,
        month = aug,
       volume = {478},
       number = {4},
        pages = {4851-4865},
          doi = {10.1093/mnras/sty1378},
archivePrefix = {arXiv},
       eprint = {1805.09865},
 primaryClass = {astro-ph.GA},
       adsurl = {https://ui.adsabs.harvard.edu/abs/2018MNRAS.478.4851I}
}

@ARTICLE{2019ApJ_885_57W,
       author = {{Wang}, Bingjie and {Heckman}, Timothy M. and {Leitherer}, Claus and {Alexandroff}, Rachel and {Borthakur}, Sanchayeeta and {Overzier}, Roderik A.},
        title = "{A New Technique for Finding Galaxies Leaking Lyman-continuum Radiation: [S II]-deficiency}",
      journal = {\apj},
         year = 2019,
        month = nov,
       volume = {885},
       number = {1},
          eid = {57},
        pages = {57},
          doi = {10.3847/1538-4357/ab418f},
archivePrefix = {arXiv},
       eprint = {1909.01368},
 primaryClass = {astro-ph.GA},
       adsurl = {https://ui.adsabs.harvard.edu/abs/2019ApJ...885...57W}
}

@ARTICLE{2021MNRAS_503_1734I,
       author = {{Izotov}, Y.~I. and {Worseck}, G. and {Schaerer}, D. and {Guseva}, N.~G. and {Chisholm}, J. and {Thuan}, T.~X. and {Fricke}, K.~J. and {Verhamme}, A.},
        title = "{Lyman continuum leakage from low-mass galaxies with M$_{{\ensuremath{\star}}}$ < {}10$^{8}$ M$_{{\ensuremath{\odot}}}$}",
      journal = {\mnras},
         year = 2021,
        month = may,
       volume = {503},
       number = {2},
        pages = {1734-1752},
          doi = {10.1093/mnras/stab612},
archivePrefix = {arXiv},
       eprint = {2103.01514},
 primaryClass = {astro-ph.GA},
       adsurl = {https://ui.adsabs.harvard.edu/abs/2021MNRAS.503.1734I}
}

@ARTICLE{2013ApJ_766_91J,
       author = {{Jaskot}, A.~E. and {Oey}, M.~S.},
        title = "{The Origin and Optical Depth of Ionizing Radiation in the ``Green Pea'' Galaxies}",
      journal = {\apj},
         year = 2013,
        month = apr,
       volume = {766},
       number = {2},
          eid = {91},
        pages = {91},
          doi = {10.1088/0004-637X/766/2/91},
archivePrefix = {arXiv},
       eprint = {1301.0530},
 primaryClass = {astro-ph.CO},
       adsurl = {https://ui.adsabs.harvard.edu/abs/2013ApJ...766...91J}
}

@ARTICLE{2014MNRAS_442_900N,
       author = {{Nakajima}, Kimihiko and {Ouchi}, Masami},
        title = "{Ionization state of inter-stellar medium in galaxies: evolution, SFR-M$_{*}$-Z dependence, and ionizing photon escape}",
      journal = {\mnras},
         year = 2014,
        month = jul,
       volume = {442},
       number = {1},
        pages = {900-916},
          doi = {10.1093/mnras/stu902},
archivePrefix = {arXiv},
       eprint = {1309.0207},
 primaryClass = {astro-ph.CO},
       adsurl = {https://ui.adsabs.harvard.edu/abs/2014MNRAS.442..900N}
}

@ARTICLE{2022AA_663A_59S,
       author = {{Saldana-Lopez}, Alberto and {Schaerer}, Daniel and {Chisholm}, John and {Flury}, Sophia R. and {Jaskot}, Anne E. and {Worseck}, G{\'a}bor and {Makan}, Kirill and {Gazagnes}, Simon and {Mauerhofer}, Valentin and {Verhamme}, Anne and {Amor{\'\i}n}, Ricardo O. and {Ferguson}, Harry C. and {Giavalisco}, Mauro and {Grazian}, Andrea and {Hayes}, Matthew J. and {Heckman}, Timothy M. and {Henry}, Alaina and {Ji}, Zhiyuan and {Marques-Chaves}, Rui and {McCandliss}, Stephan R. and {Oey}, M. Sally and {{\"O}stlin}, G{\"o}ran and {Pentericci}, Laura and {Thuan}, Trinh X. and {Trebitsch}, Maxime and {Vanzella}, Eros and {Xu}, Xinfeng},
        title = "{The Low-Redshift Lyman Continuum Survey. Unveiling the ISM properties of low-z Lyman-continuum emitters}",
      journal = {\aap},
         year = 2022,
        month = jul,
       volume = {663},
          eid = {A59},
        pages = {A59},
          doi = {10.1051/0004-6361/202141864},
archivePrefix = {arXiv},
       eprint = {2201.11800},
 primaryClass = {astro-ph.GA},
       adsurl = {https://ui.adsabs.harvard.edu/abs/2022A&A...663A..59S}
}

@ARTICLE{2022MNRAS_517_5104C,
       author = {{Chisholm}, J. and {Saldana-Lopez}, A. and {Flury}, S. and {Schaerer}, D. and {Jaskot}, A. and {Amor{\'\i}n}, R. and {Atek}, H. and {Finkelstein}, S.~L. and {Fleming}, B. and {Ferguson}, H. and {Fern{\'a}ndez}, V. and {Giavalisco}, M. and {Hayes}, M. and {Heckman}, T. and {Henry}, A. and {Ji}, Z. and {Marques-Chaves}, R. and {Mauerhofer}, V. and {McCandliss}, S. and {Oey}, M.~S. and {{\"O}stlin}, G. and {Rutkowski}, M. and {Scarlata}, C. and {Thuan}, T. and {Trebitsch}, M. and {Wang}, B. and {Worseck}, G. and {Xu}, X.},
        title = "{The far-ultraviolet continuum slope as a Lyman Continuum escape estimator at high redshift}",
      journal = {\mnras},
         year = 2022,
        month = dec,
       volume = {517},
       number = {4},
        pages = {5104-5120},
          doi = {10.1093/mnras/stac2874},
archivePrefix = {arXiv},
       eprint = {2207.05771},
 primaryClass = {astro-ph.GA},
       adsurl = {https://ui.adsabs.harvard.edu/abs/2022MNRAS.517.5104C}
}

@INCOLLECTION{Mange1972-yl,
  title     = "The Exosphere and Geocorona",
  booktitle = "The Upper Atmosphere",
  author    = "Mange, P",
  publisher = "Springer Netherlands",
  pages     = "68--86",
  year      =  1972,
  address   = "Dordrecht"
}

@ARTICLE{2010ApJ_709L_183M,
       author = {{McCandliss}, Stephan R. and {France}, Kevin and {Osterman}, Steven and {Green}, James C. and {McPhate}, Jason B. and {Wilkinson}, Erik},
        title = "{Far-Ultraviolet Sensitivity of the Cosmic Origins Spectrograph}",
      journal = {\apjl},
         year = 2010,
        month = feb,
       volume = {709},
       number = {2},
        pages = {L183-L187},
          doi = {10.1088/2041-8205/709/2/L183},
archivePrefix = {arXiv},
       eprint = {0909.3878},
 primaryClass = {astro-ph.IM},
       adsurl = {https://ui.adsabs.harvard.edu/abs/2010ApJ...709L.183M}
}

@ARTICLE{2008MNRAS_387_1681I,
       author = {{Inoue}, Akio K. and {Iwata}, Ikuru},
        title = "{A Monte Carlo simulation of the intergalactic absorption and the detectability of the Lyman continuum from distant galaxies}",
      journal = {\mnras},
         year = 2008,
        month = jul,
       volume = {387},
       number = {4},
        pages = {1681-1692},
          doi = {10.1111/j.1365-2966.2008.13350.x},
archivePrefix = {arXiv},
       eprint = {0804.2951},
 primaryClass = {astro-ph},
       adsurl = {https://ui.adsabs.harvard.edu/abs/2008MNRAS.387.1681I}
}

@ARTICLE{2024Natur_633_318I,
	author = {Carniani, Stefano and Hainline, Kevin and D'Eugenio, Francesco and Eisenstein, Daniel J. and Jakobsen, Peter and Witstok, Joris and Johnson, Benjamin D. and Chevallard, Jacopo and Maiolino, Roberto and Helton, Jakob M. and Willott, Chris and Robertson, Brant and Alberts, Stacey and Arribas, Santiago and Baker, William M. and Bhatawdekar, Rachana and Boyett, Kristan and Bunker, Andrew J. and Cameron, Alex J. and Cargile, Phillip A. and Charlot, St{\'e}phane and Curti, Mirko and Curtis-Lake, Emma and Egami, Eiichi and Giardino, Giovanna and Isaak, Kate and Ji, Zhiyuan and Jones, Gareth C. and Kumari, Nimisha and Maseda, Michael V. and Parlanti, Eleonora and P{\'e}rez-Gonz{\'a}lez, Pablo G. and Rawle, Tim and Rieke, George and Rieke, Marcia and Del Pino, Bruno Rodr{\'\i}guez and Saxena, Aayush and Scholtz, Jan and Smit, Renske and Sun, Fengwu and Tacchella, Sandro and {\"U}bler, Hannah and Venturi, Giacomo and Williams, Christina C. and Willmer, Christopher N. A.},
	journal = {Nature},
	number = {8029},
	pages = {318--322},
	title = {Spectroscopic confirmation of two luminous galaxies at a redshift of 14},
	volume = {633},
	year = {2024}}

@ARTICLE{2023NatAs_7_611R,
       author = {{Robertson}, B.~E. and {Tacchella}, S. and {Johnson}, B.~D. and {Hainline}, K. and {Whitler}, L. and {Eisenstein}, D.~J. and {Endsley}, R. and {Rieke}, M. and {Stark}, D.~P. and {Alberts}, S. and {Dressler}, A. and {Egami}, E. and {Hausen}, R. and {Rieke}, G. and {Shivaei}, I. and {Williams}, C.~C. and {Willmer}, C.~N.~A. and {Arribas}, S. and {Bonaventura}, N. and {Bunker}, A. and {Cameron}, A.~J. and {Carniani}, S. and {Charlot}, S. and {Chevallard}, J. and {Curti}, M. and {Curtis-Lake}, E. and {D'Eugenio}, F. and {Jakobsen}, P. and {Looser}, T.~J. and {L{\"u}tzgendorf}, N. and {Maiolino}, R. and {Maseda}, M.~V. and {Rawle}, T. and {Rix}, H. -W. and {Smit}, R. and {{\"U}bler}, H. and {Willott}, C. and {Witstok}, J. and {Baum}, S. and {Bhatawdekar}, R. and {Boyett}, K. and {Chen}, Z. and {de Graaff}, A. and {Florian}, M. and {Helton}, J.~M. and {Hviding}, R.~E. and {Ji}, Z. and {Kumari}, N. and {Lyu}, J. and {Nelson}, E. and {Sandles}, L. and {Saxena}, A. and {Suess}, K.~A. and {Sun}, F. and {Topping}, M. and {Wallace}, I.~E.~B.},
        title = "{Identification and properties of intense star-forming galaxies at redshifts z > 10}",
      journal = {Nature Astronomy},
         year = 2023,
        month = may,
       volume = {7},
        pages = {611-621},
          doi = {10.1038/s41550-023-01921-1},
archivePrefix = {arXiv},
       eprint = {2212.04480},
 primaryClass = {astro-ph.GA},
       adsurl = {https://ui.adsabs.harvard.edu/abs/2023NatAs...7..611R}
}

@ARTICLE{2023ApJ_957L_34W,
       author = {{Wang}, Bingjie and {Fujimoto}, Seiji and {Labb{\'e}}, Ivo and {Furtak}, Lukas J. and {Miller}, Tim B. and {Setton}, David J. and {Zitrin}, Adi and {Atek}, Hakim and {Bezanson}, Rachel and {Brammer}, Gabriel and {Leja}, Joel and {Oesch}, Pascal A. and {Price}, Sedona H. and {Chemerynska}, Iryna and {Cutler}, Sam E. and {Dayal}, Pratika and {van Dokkum}, Pieter and {Goulding}, Andy D. and {Greene}, Jenny E. and {Fudamoto}, Y. and {Khullar}, Gourav and {Kokorev}, Vasily and {Marchesini}, Danilo and {Pan}, Richard and {Weaver}, John R. and {Whitaker}, Katherine E. and {Williams}, Christina C.},
        title = "{UNCOVER: Illuminating the Early Universe-JWST/NIRSpec Confirmation of z > 12 Galaxies}",
      journal = {\apjl},
         year = 2023,
        month = nov,
       volume = {957},
       number = {2},
          eid = {L34},
        pages = {L34},
          doi = {10.3847/2041-8213/acfe07},
archivePrefix = {arXiv},
       eprint = {2308.03745},
 primaryClass = {astro-ph.GA},
       adsurl = {https://ui.adsabs.harvard.edu/abs/2023ApJ...957L..34W}
}

@INPROCEEDINGS{2019SPIE11118E_0UF,
       author = {{Fleming}, Brian T. and {France}, Kevin and {Williams}, Jack and {Ulrich}, Stefan and {Tumlinson}, Jason and {McCandliss}, Stephan and {O'Meara}, John and {Sankrit}, Ravi and {Borthakur}, Sanchayeeta and {Jaskot}, Anne and {Rutkowski}, Michael and {Quijada}, Manuel and {Hennessy}, John and {Siegmund}, Oswald},
        title = "{High-sensitivity far-ultraviolet imaging spectroscopy with the SPRITE Cubesat}",
    booktitle = {UV, X-Ray, and Gamma-Ray Space Instrumentation for Astronomy XXI},
         year = 2019,
       editor = {{Siegmund}, Oswald H.},
       series = {Society of Photo-Optical Instrumentation Engineers (SPIE) Conference Series},
       volume = {11118},
        month = sep,
          eid = {111180U},
        pages = {111180U},
          doi = {10.1117/12.2529512},
       adsurl = {https://ui.adsabs.harvard.edu/abs/2019SPIE11118E..0UF}
}

@ARTICLE{2022NatAs_6_1213F,
       author = {{Fleming}, Brian},
        title = "{Mapping the influence of massive stars}",
      journal = {Nature Astronomy},
         year = 2022,
        month = oct,
       volume = {6},
        pages = {1213-1213},
          doi = {10.1038/s41550-022-01807-8},
       adsurl = {https://ui.adsabs.harvard.edu/abs/2022NatAs...6.1213F}
}

@INPROCEEDINGS{2023SPIE12678E_06I,
       author = {{Indahl}, Briana and {Fleming}, Brian and {Vorobiev}, Dmitry and {Chafetz}, Dana and {Williams}, Jack and {Bowen}, Maitland and {Brening}, Diane and {Borthakur}, Sanchayeeta and {Del Hoyo}, Javier and {Dewitt}, Destry and {Diaz}, Adriana and {Durell}, Abigail and {Foehr}, Ben and {France}, Kevin and {Gopinathan}, Sreejith and {Hennessy}, John and {Jaskot}, Anne and {Kaiser}, Michael and {Koehler}, Sydney and {Magruder}, Adam and {Martin}, Adrian and {McCandliss}, Stephan and {O'Meara}, John and {Quijada}, Manuel and {Rodr{\'\i}guez-de Marcos}, Luis and {Rotkowski}, Michael and {Sankrit}, Ravi and {Sico}, Alex and {Siegmund}, Oswald H. and {Szewczyk}, Daniel and {Tumlinson}, Jason and {Ulrich}, Stefan},
        title = "{Status and mission operations of the SPRITE 12U CubeSat: a probe of star formation feedback from stellar to galactic scales with far-UV imaging spectroscopy}",
    booktitle = {UV, X-Ray, and Gamma-Ray Space Instrumentation for Astronomy XXIII},
         year = 2023,
       editor = {{Siegmund}, Oswald H. and {Hoadley}, Keri},
       series = {Society of Photo-Optical Instrumentation Engineers (SPIE) Conference Series},
       volume = {12678},
        month = oct,
          eid = {1267806},
        pages = {1267806},
          doi = {10.1117/12.2677737},
       adsurl = {https://ui.adsabs.harvard.edu/abs/2023SPIE12678E..06I}
}

@inproceedings{2024SPIE1309332,
        author = {Donal O'Sullivan and Brian Fleming and Briana Indahl and Dmitry   Vorobiev and Maitland Bowen and Kevin France and Sebastian Escobar and Yi Hang Valerie Wong and Elena Carlson and Sreejith Aickara Gopinathan and Sanchayeeta Borthakur and Javier Del Hoyo and Adriana Diaz and John Hennessy and Anne Jaskot and Adam Magruder and Adrian Martin and Stephan McCandliss and John O'Meara and Manuel Quijada and Luis Rodr{\'i}guez-de Marcos and Michael Rutkowski and Ravi Sankrit and Oswald H. Siegmund and Jason Tumlinson and Dana Chafetz and Stefan Ulrich and Beth Cervelli and Jack Williams and Diane Brening and Alex Sico and Michael Kaiser},
    title = {{Observing modes of the SPRITE 12U CubeSat: a probe of star formation feedback with far-UV imaging spectroscopy}},
    volume = {13093},
    booktitle = {Space Telescopes and Instrumentation 2024: Ultraviolet to Gamma Ray},
    editor = {Jan-Willem A. den Herder and Shouleh Nikzad and Kazuhiro Nakazawa},
    organization = {International Society for Optics and Photonics},
    publisher = {SPIE},
    pages = {1309332},
    year = {2024},
    doi = {10.1117/12.3020332},
    URL = {https://doi.org/10.1117/12.3020332}
}

@ARTICLE{2017ApOpt_56_9941F,
       author = {{Fleming}, Brian and {Quijada}, Manuel and {Hennessy}, John and {Egan}, Arika and {Del Hoyo}, Javier and {Hicks}, Brian A. and {Wiley}, James and {Kruczek}, Nicholas and {Erickson}, Nicholas and {France}, Kevin},
        title = "{Advanced environmentally resistant lithium fluoride mirror coatings for the next generation of broadband space observatories}",
      journal = {\ao},
         year = 2017,
        month = dec,
       volume = {56},
       number = {36},
        pages = {9941},
          doi = {10.1364/ao.56.009941},
       adsurl = {https://ui.adsabs.harvard.edu/abs/2017ApOpt..56.9941F}
}

@INPROCEEDINGS{2022SPIE12188E_20R,
       author = {{Rodr{\'\i}guez-de Marcos}, L. and {Fleming}, B. and {Hennessy}, J. and {Chafetz}, D. and {Del Hoyo}, J. and {Quijada}, M. and {Bowen}, M. and {Vorobiev}, D. and {Indahl}, B.},
        title = "{Advanced Al/eLiF mirrors for the SPRITE CubeSat}",
    booktitle = {Advances in Optical and Mechanical Technologies for Telescopes and Instrumentation},
         year = 2022,
       series = {Society of Photo-Optical Instrumentation Engineers (SPIE) Conference Series},
       volume = {12188},
        month = aug,
          eid = {1218820},
        pages = {1218820},
          doi = {10.1117/12.2630522},
       adsurl = {https://ui.adsabs.harvard.edu/abs/2022SPIE12188E..20R}
}

@INPROCEEDINGS{2022SPIE12188E_1VQ,
       author = {{Quijada}, Manuel A. and {Rodriguez de Marcos}, Luis V. and {Del Hoyo}, Javier G. and {Gray}, Emrold and {Wollack}, Edward J. and {Brown}, Ari},
        title = "{Advanced Al mirrors protected with LiF overcoat to realize stable mirror coatings for astronomical telescopes}",
    booktitle = {Advances in Optical and Mechanical Technologies for Telescopes and Instrumentation},
         year = 2022,
       series = {Society of Photo-Optical Instrumentation Engineers (SPIE) Conference Series},
       volume = {12188},
        month = aug,
          eid = {121881V},
        pages = {121881V},
          doi = {10.1117/12.2630585},
       adsurl = {https://ui.adsabs.harvard.edu/abs/2022SPIE12188E..1VQ}
}

@article{carlson2025push,
  title={Push-broom mapping of galaxies and supernova remnants with the SPRITE CubeSat},
  author={Carlson, Elena and Fleming, Brian and Valerie Wong, Yi Hang and Indahl, Briana and Vorobiev, Dmitry and Bowen, Maitland and O’Sullivan, Donal and France, Kevin and Jaskot, Anne and Tumlinson, Jason and others},
  journal={Journal of Astronomical Telescopes, Instruments, and Systems},
  volume={11},
  number={4},
  pages={045001--045001},
  year={2025},
  publisher={Society of Photo-Optical Instrumentation Engineers}
}

@INPROCEEDINGS{2020SPIE11454E_1HS,
       author = {{Siegmund}, O.~H.~W. and {McPhate}, J.~B. and {Curtis}, T. and {Darling}, N. and {Vallerga}, J.~V. and {Cremer}, T. and {Ertley}, C.},
        title = "{Development of UV imaging detectors with atomic layer deposited microchannel plates and cross strip readouts}",
    booktitle = {X-Ray, Optical, and Infrared Detectors for Astronomy IX},
         year = 2020,
       editor = {{Holland}, Andrew D. and {Beletic}, James},
       series = {Society of Photo-Optical Instrumentation Engineers (SPIE) Conference Series},
       volume = {11454},
        month = dec,
          eid = {114541H},
        pages = {114541H},
          doi = {10.1117/12.2561753},
       adsurl = {https://ui.adsabs.harvard.edu/abs/2020SPIE11454E..1HS}
}

@INPROCEEDINGS{2021SPIE11821E_0HC,
       author = {{Cruz Aguirre}, Fernando and {Nell}, Nicholas and {Kruczek}, Nicholas and {Hinton}, P.~C. and {Bridges}, Matthew and {France}, Kevin and {Fleming}, Brian},
        title = "{The assembly, calibration, and predicted performance of the SISTINE-2 sounding rocket payload}",
    booktitle = {UV, X-Ray, and Gamma-Ray Space Instrumentation for Astronomy XXII},
         year = 2021,
       editor = {{Siegmund}, Oswald H.},
       series = {Society of Photo-Optical Instrumentation Engineers (SPIE) Conference Series},
       volume = {11821},
        month = aug,
          eid = {118210H},
        pages = {118210H},
          doi = {10.1117/12.2594393},
       adsurl = {https://ui.adsabs.harvard.edu/abs/2021SPIE11821E..0HC}
}

@INPROCEEDINGS{2000SPIE_4013_360O,
       author = {{Osterman}, Steven N. and {Wilkinson}, Erik and {Green}, James C. and {Redman}, Kevin W.},
        title = "{FUV grating performance for the cosmic origins spectrograph}",
    booktitle = {UV, Optical, and IR Space Telescopes and Instruments},
         year = 2000,
       editor = {{Breckinridge}, James B. and {Jakobsen}, Peter},
       series = {Society of Photo-Optical Instrumentation Engineers (SPIE) Conference Series},
       volume = {4013},
        month = jul,
        pages = {360-366},
          doi = {10.1117/12.394019},
       adsurl = {https://ui.adsabs.harvard.edu/abs/2000SPIE.4013..360O}
}

@ARTICLE{2024MNRAS_530_2133S,
       author = {{Simmonds}, C. and {Verhamme}, A. and {Inoue}, A.~K. and {Katz}, H. and {Garel}, T. and {De Barros}, S.},
        title = "{The impact of nebular Lyman-Continuum on ionizing photons budget and escape fractions from galaxies}",
      journal = {\mnras},
         year = 2024,
        month = may,
       volume = {530},
       number = {2},
        pages = {2133-2145},
          doi = {10.1093/mnras/stae1003},
archivePrefix = {arXiv},
       eprint = {2402.04052},
 primaryClass = {astro-ph.GA},
       adsurl = {https://ui.adsabs.harvard.edu/abs/2024MNRAS.530.2133S}
}

@INPROCEEDINGS{2014SPIE_9144E_4HM,
       author = {{Moore}, Christopher S. and {Hennessy}, John and {Jewell}, April D. and {Nikzad}, Shouleh and {France}, Kevin},
        title = "{Recent developments and results of new ultraviolet reflective mirror coatings}",
    booktitle = {Space Telescopes and Instrumentation 2014: Ultraviolet to Gamma Ray},
         year = 2014,
       editor = {{Takahashi}, Tadayuki and {den Herder}, Jan-Willem A. and {Bautz}, Mark},
       series = {Society of Photo-Optical Instrumentation Engineers (SPIE) Conference Series},
       volume = {9144},
        month = jul,
          eid = {91444H},
        pages = {91444H},
          doi = {10.1117/12.2057272},
       adsurl = {https://ui.adsabs.harvard.edu/abs/2014SPIE.9144E..4HM}
}

@INPROCEEDINGS{1996SPIE_2807_172K,
       author = {{Kennedy}, Michael J. and {Friedman}, Scott D. and {Barkhouser}, Robert H. and {Hampton}, Jeff and {Nikulla}, Paul},
        title = "{Design of the Far Ultraviolet Spectroscopic Explorer mirror assemblies}",
    booktitle = {Space Telescopes and Instruments IV},
         year = 1996,
       editor = {{Bely}, Pierre Y. and {Breckinridge}, James B.},
       series = {Society of Photo-Optical Instrumentation Engineers (SPIE) Conference Series},
       volume = {2807},
        month = oct,
        pages = {172-183},
          doi = {10.1117/12.255098},
       adsurl = {https://ui.adsabs.harvard.edu/abs/1996SPIE.2807..172K}
}

@ARTICLE{2016PASP_128j5006R,
       author = {{Redwine}, Keith and {McCandliss}, Stephan R. and {Zheng}, Wei and {Fleming}, Brian and {France}, Kevin and {Osterman}, Steven and {Howk}, J. Christopher and {Anderson}, Scott F. and {G{\"a}ensicke}, Boris T.},
        title = "{New Gapless COS G140L Mode Proposed for Background-limited Far-UV Observations}",
      journal = {\pasp},
         year = 2016,
        month = oct,
       volume = {128},
       number = {968},
        pages = {105006},
          doi = {10.1088/1538-3873/128/968/105006},
archivePrefix = {arXiv},
       eprint = {1606.00402},
 primaryClass = {astro-ph.IM},
       adsurl = {https://ui.adsabs.harvard.edu/abs/2016PASP..128j5006R}
}

@ARTICLE{2001ApJ_546_665S,
       author = {{Steidel}, Charles C. and {Pettini}, Max and {Adelberger}, Kurt L.},
        title = "{Lyman-Continuum Emission from Galaxies at Z \raisebox{-0.5ex}\textasciitilde= 3.4}",
      journal = {\apj},
         year = 2001,
        month = jan,
       volume = {546},
       number = {2},
        pages = {665-671},
          doi = {10.1086/318323},
archivePrefix = {arXiv},
       eprint = {astro-ph/0008283},
 primaryClass = {astro-ph},
       adsurl = {https://ui.adsabs.harvard.edu/abs/2001ApJ...546..665S}
}

@ARTICLE{2019ApJ_882_182C,
       author = {{Chisholm}, J. and {Rigby}, J.~R. and {Bayliss}, M. and {Berg}, D.~A. and {Dahle}, H. and {Gladders}, M. and {Sharon}, K.},
        title = "{Constraining the Metallicities, Ages, Star Formation Histories, and Ionizing Continua of Extragalactic Massive Star Populations}",
      journal = {\apj},
         year = 2019,
        month = sep,
       volume = {882},
       number = {2},
          eid = {182},
        pages = {182},
          doi = {10.3847/1538-4357/ab3104},
archivePrefix = {arXiv},
       eprint = {1905.04314},
 primaryClass = {astro-ph.GA},
       adsurl = {https://ui.adsabs.harvard.edu/abs/2019ApJ...882..182C}
}

@ARTICLE{2023AJ_165_64E,
       author = {{Egan}, Arika and {Nell}, Nicholas and {Suresh}, Ambily and {France}, Kevin and {Fleming}, Brian and {Sreejith}, Aickara Gopinathan and {Lambert}, Julian and {DeCicco}, Nicholas},
        title = "{The On-orbit Performance of the Colorado Ultraviolet Transit Experiment Mission}",
      journal = {\aj},
         year = 2023,
        month = feb,
       volume = {165},
       number = {2},
          eid = {64},
        pages = {64},
          doi = {10.3847/1538-3881/aca8a3},
archivePrefix = {arXiv},
       eprint = {2301.01307},
 primaryClass = {astro-ph.IM},
       adsurl = {https://ui.adsabs.harvard.edu/abs/2023AJ....165...64E}
}

@ARTICLE{2023AJ_165_63F,
       author = {{France}, Kevin and {Fleming}, Brian and {Egan}, Arika and {Desert}, Jean-Michel and {Fossati}, Luca and {Koskinen}, Tommi T. and {Nell}, Nicholas and {Petit}, Pascal and {Vidotto}, Aline A. and {Beasley}, Matthew and {DeCicco}, Nicholas and {Sreejith}, Aickara Gopinathan and {Suresh}, Ambily and {Baumert}, Jared and {Cauley}, P. Wilson and {Villarreal D'Angelo}, Carolina and {Hoadley}, Keri and {Kane}, Robert and {Kohnert}, Richard and {Lambert}, Julian and {Ulrich}, Stefan},
        title = "{The Colorado Ultraviolet Transit Experiment Mission Overview}",
      journal = {\aj},
         year = 2023,
        month = feb,
       volume = {165},
       number = {2},
          eid = {63},
        pages = {63},
          doi = {10.3847/1538-3881/aca8a2},
archivePrefix = {arXiv},
       eprint = {2301.02250},
 primaryClass = {astro-ph.IM},
       adsurl = {https://ui.adsabs.harvard.edu/abs/2023AJ....165...63F}
}

@article{JATIS_11_4_042223,
author = {Stephan Robert McCandliss},
title = {{Direct detection of ionizing radiation with Habitable Worlds Observatory}},
volume = {11},
journal = {Journal of Astronomical Telescopes, Instruments, and Systems},
number = {4},
publisher = {SPIE},
pages = {042223},
year = {2025},
doi = {10.1117/1.JATIS.11.4.042223},
URL = {https://doi.org/10.1117/1.JATIS.11.4.042223}
}

@ARTICLE{2017ApJ_847_38Y,
       author = {{Yang}, Huan and {Malhotra}, Sangeeta and {Rhoads}, James E. and {Wang}, Junxian},
        title = "{Blueberry Galaxies: The Lowest Mass Young Starbursts}",
      journal = {\apj},
         year = 2017,
        month = sep,
       volume = {847},
       number = {1},
          eid = {38},
        pages = {38},
          doi = {10.3847/1538-4357/aa8809},
archivePrefix = {arXiv},
       eprint = {1706.02819},
 primaryClass = {astro-ph.GA},
       adsurl = {https://ui.adsabs.harvard.edu/abs/2017ApJ...847...38Y}
}

@ARTICLE{2016JAI_540001F,
       author = {{France}, Kevin and {Hoadley}, Keri and {Fleming}, Brian T. and {Kane}, Robert and {Nell}, Nicholas and {Beasley}, Matthew and {Green}, James C.},
        title = "{The SLICE, CHESS, and SISTINE Ultraviolet Spectrographs: Rocket-Borne Instrumentation Supporting Future Astrophysics Missions}",
      journal = {Journal of Astronomical Instrumentation},
         year = 2016,
        month = dec,
       volume = {5},
       number = {1},
          eid = {1640001},
        pages = {1640001},
          doi = {10.1142/S2251171716400018},
archivePrefix = {arXiv},
       eprint = {1512.00881},
 primaryClass = {astro-ph.IM},
       adsurl = {https://ui.adsabs.harvard.edu/abs/2016JAI.....540001F}
}

@ARTICLE{2025JATIS_11d2209Q,
       author = {{Quijada}, Manuel A. and {Del Hoyo}, Javier G. and {Rodriguez de Marcos}, Luis V. and {Wollack}, Edward J. and {Batkis}, Mateo F. and {Lewis}, Devin M. and {Rydalch}, Tanner D. and {Allred}, David D.},
        title = "{Far-ultraviolet optical properties and performance of physical vapor deposited aluminum mirrors protected with XeLiF and XeMgF$_{2}$}",
      journal = {Journal of Astronomical Telescopes, Instruments, and Systems},
         year = 2025,
        month = oct,
       volume = {11},
          eid = {042209},
        pages = {042209},
          doi = {10.1117/1.JATIS.11.4.042209},
       adsurl = {https://ui.adsabs.harvard.edu/abs/2025JATIS..11d2209Q}
}
\bibliographystyle{spiejour}   

\vspace{2ex}\noindent\textbf{Yi Hang Valerie Wong} is a doctoral researcher at LASP/CU Boulder, focusing on data reduction and analysis for SPRITE and understanding the star formation and galaxy evolution through ultraviolet and optical observations.

\vspace{2ex}\noindent\textbf{Elena Carlson} is an undergraduate research assistant at the Laboratory for Atmospheric and Space Physics, University of Colorado Boulder. She works on the SPRITE SmallSat mission, focusing on SmallSat instrumentation and spectral modeling of supernova remnants in the far-UV. Her research interests include astrophysics, spacecraft instrumentation and design, and remote sensing technologies. She plans to pursue graduate study in aerospace engineering.

\vspace{2ex}\noindent\textbf{Kevin France} is a professor at LASP/University of Colorado. His research focuses on exoplanets, their host stars, and the development of instrumentation for ultraviolet astrophysics. He is the PI of the ESCAPE Small Explorer concept, NASA’s CUTE mission, and a NASA-supported sounding rocket to flight-test critical path hardware for future UV/optical astrophysics missions. He was a member of the \textit{HST}-COS and HWO START teams, and the study PI for the LUVOIR ultraviolet spectrograph.

\vspace{2ex}\noindent\textbf{Anne Jaskot} earned her PhD in astrophysics at the University of Michigan and is now an associate professor at Williams College. She uses ultraviolet spectroscopy to study the production and escape of ionizing radiation from low-redshift starburst galaxies.

\vspace{2ex}\noindent\textbf{Sanchayeeta Borthakur} is an associate professor at Arizona State University. She received her PhD in astronomy from the University of Massachusetts Amherst in 2010. She has authored/co-authored more than 50 journal articles. Her research interests include UV spectroscopy, studies of the CGM, the IGM, and the ISM. She has led multiple programs with the \textit{Hubble Space Telescope} and the Very Large Array, and is the co-chair of the IGM/CGM science working group for the Habitable Worlds Observatory.

\vspace{2ex}\noindent\textbf{Michael Rutkowski} is an associate professor at Minnesota State University in the Department of Physics and Astronomy. His research and teaching is focused on extragalactic astrophysics of massive, quiescent galaxy formation and evolution and the impact of star formation on the CGM/IGM using space-based observatories like \textit{Hubble}, \textit{GALEX}, AstroSat, and \textit{JWST}. Prior to Minnesota State, he was a post-doctoral research associate at Stockholm University and the University of Minnesota; he completed his PhD in 2013.

\vspace{2ex}\noindent\textbf{John M. O'Meara} is the chief scientist and deputy director of the W. M. Keck Observatory. He served as the co-chair of the Habitable Worlds Observatory START team and as the Cosmic Origins Co-Lead for the LUVOIR large mission concept study. He has served as the chair/co-chair of the astrophysics senior review, and currently serves as co-chair of the Committee on Astrophysics and Astronomy for the National Academies.

\vspace{1ex}
\noindent Biographies and photographs of the other authors are not available.

\listoffigures
\listoftables

\end{spacing}
\end{document}